\documentclass[twocolumn]{aastex63}
\usepackage{soul}
\usepackage[flushleft]{threeparttable}
\usepackage[utf8]{inputenc}
\usepackage{subfigure} 
\usepackage{float}
\usepackage{multirow}
\usepackage{graphicx}
\usepackage{amssymb}
\usepackage{amsmath}
\usepackage{hyperref}
\usepackage{url}
\usepackage{fancyhdr}

\def\simlt{\mathrel{\rlap{\lower 3pt\hbox{$\sim$}}
        \raise 2.0pt\hbox{$<$}}}
\def\simgt{\mathrel{\rlap{\lower 3pt\hbox{$\sim$}}
        \raise 2.0pt\hbox{$>$}}}

\revised{\today}

\submitjournal{ApJ}

\begin{document}

\def\jdg#1{{\bf[#1 -- JDG]}}
\def\JA#1{{\bf[#1 -- JA]}}

\title{Determining the nature of IC 10 X-2: A comprehensive study of the OIR emission from an extragalactic BeHMXB} 

\author[0000-0002-4644-6580]{Jwaher Alnaqbi}
\affiliation{Center for Astrophysics and Space Science (CASS), New York University Abu Dhabi, PO Box 129188, Abu Dhabi, UAE }

\author[0000-0003-4679-1058]{Joseph D. Gelfand}
\affiliation{Center for Astrophysics and Space Science (CASS), New York University Abu Dhabi, PO Box 129188, Abu Dhabi, UAE }
\affiliation{Affiliate Member, Center for Cosmology and Particle Physics (CCPP), New York University, 726 Broadway, New York, NY 10003}
\correspondingauthor USA {Joseph D. Gelfand}
\author[0000-0002-5319-6620]{Payaswini Saikia}
\affiliation{Center for Astrophysics and Space Science (CASS), New York University Abu Dhabi, PO Box 129188, Abu Dhabi, UAE }

\author[0000-0003-3944-6109]{Craig Heinke}
\affiliation{Physics Dept., CCIS 4-183, University of Alberta, Edmonton, AB T6G 2E1, Canada }

\author[0000-0003-1285-4057]{M. C. Baglio}
\affiliation{INAF, Osservatorio Astronomico di Brera, Via Bianchi 46, Merate (LC), I-23807, Italy }

\author[0000-0002-3500-631X]{David M. Russell}
\affiliation{Center for Astrophysics and Space Science (CASS), New York University Abu Dhabi, PO Box 129188, Abu Dhabi, UAE }

\author[0000-0001-8630-5435]{Guobao Zhang}
\affiliation{Yunnan Observatories Chinese Academy of Sciences}

\author{Antonios Manousakis}
\affiliation{Department of Applied Physics and Astronomy, College of Sciences, University of Sharjah POBox 27272 University City, Sharjah, UAE and Sharjah Academy for Astronomy, Space Sciences and Technology (SAASST), POBox 27272 University City, Sharjah, UAE}

\author[0000-0002-6425-6879]{Ivan Yu. Katkov}
\affiliation{Center for Astrophysics and Space Science (CASS), New York University Abu Dhabi, PO Box 129188, Abu Dhabi, UAE }
\affiliation{Sternberg Astronomical Institute, Lomonosov Moscow State University, Universitetskij pr., 13,  Moscow, 119234, Russia}

\author[0000-0003-3352-2334]{Fraser Lewis}
\affiliation{Faulkes Telescope Project, School of Physics and Astronomy, Cardiff University, The Parade, Cardiff, CF24 3AA Wales, UK}

\affiliation{Astrophysics Research Institute, Liverpool John Moores University, 146 Brownlow Hill, Liverpool L3 5RF, UK}


\begin{abstract}
We present a comprehensive analysis of the optical and infrared (IR) properties of high-mass X-ray binary (HMXB) IC 10 X-2, classified as a super-giant HMXB and super-fast X-ray transient (SFXT) by previous work.  Our analysis of regular (daily and weekly) observations by both the Zwicky Transient Factory and Las Cumbres Observatory over a ~5 year period indicates both periodic flares and variations in the apparent magnitude and color with a $\sim26.5$~d period -- likely the orbital period of this binary system.  The periodic flaring suggests the stellar companion is a Be star, with flares resulting from increased accretion onto the neutron star when it enters the stellar decretion disk.  The periodic variations in the optical/IR brightness and color likely result from orbital variations in the Hydrogen column density along the line of sight or a transient accretion disk around the neutron star.  Lastly, the numerous, short duration, episodes where IC 10 X-2 is significantly ``redder'' or ``bluer'' than normal likely result from from clumps within this system -- which can accrete onto the neutron star (causing IC 10 X-2 to appear bluer), or pass through the line of sight (causing IC 10 X-2 to appear redder).  These results substantially increase our understanding of the evolution of this source, a significant source of ionizing photons in its host galaxy IC 10, a low mass, metal-poor starburst galaxy similar in many respect to those thought to be common in the early Universe.
\end{abstract}

\section{Introduction}
\label{sec:intro}

High-mass X-ray binaries (HMXBs) are believed to be systems where a
compact object, i.e., a neutron star (NS) or a black hole (BH), are
gravitationally bound to a more massive star, typically an O- or
B-type companion, either on the main sequence phase or in a later
supergiant (SG) evolutionary stage. HMXBs are primarily classified by
the mode of accretion from the stellar companion onto the compact
object, which in turn depends on the nature of the companion. 

There are two primary modes of accretion, and hence primary classes
(e.g., \citealt{Tan_2021}) of HMXBs, the first of which is Bondi ( or
``wind-fed,'' in the case of HMXBs; \citealt{Shakura_2015}) accretion
of material ejected by a, typically SG, star (sgHMXB).  SG systems
often have relatively short orbital periods ($\lesssim20$ days; e.g.,
\citealt{Townsend_2011}) and, during quiescence, the dominant source
of the optical emission is the SG star due to its high luminosity.  In
these systems, absorbing clumps are frequently detected in optical
observations as decreases in brightness, with larger amplitude
decreases observed at shorter wavelengths.   Flares originate from
clumps accreting onto the compact object, and are observed primarily
at infrared (IR), optical, ultraviolet (UV), and X-ray wavebands.  

In the second class, the stellar companion typically an Oe/Be star,
and hence systems of this type are referred to as BeHMXBs. In these
systems, the compact object \citep[generally a NS,][]{Shi_2015}
primarily accretes material from the decretion disk created by the
highly asymmetric wind of the companion star, infrequently
supplemented by minor Roche lobe overflow from the star onto the
compact object (e.g., \citealt{chaty_2018}). The orbital passage of
the NS through the circumstellar disk results in periodic outbursts or
flares with X-ray luminosities up to $\sim10^{38}$ erg/s that can last
for days or weeks \cite[]{Ducci_2022} in this and other (IR, optical,
and UV) wavebands (e.g., \citealt{Alcock_2001,Bird_2012}).  

Be systems can be identified through these periodic flares or
outbursts, that are frequently used to determine the orbital period in
these systems \citep[e.g.,][]{Alcock_2001,Jain_2009}.  Be systems can
also be distinguished by high extinction or reddening caused by the
material in the disk, and strong H$\alpha$ emission, in addition to
appearing redder when fainter \citep[e.g.,][]{Alcock_2001}. While the
optical emission from Be systems is generally dominated by the
companion star, accretion onto the compact object can also
significantly contribute during periodic flares or outbursts.   

Some Be and sgHMXBs are observed to be Supergiant Fast X-ray
Transients (SFXTs), i.e. system that exhibit short-duration, intense
X-ray flares detected at any orbital phase -- attributed to the
accretion of relatively distant clumps formed in the stellar wind of
the companion star onto the compact object
\citep{Ducci_2009,Oskinova_2007,walter_2007,bandyopadhyay_2008}. They
can also exhibit short ($\sim$a few hours long) periods of significant
X-ray (and optical) variability \cite[]{Pellizza_2006} as a result of
clumps wandering through the line of sight and blocking some of the
emission, leading dips in its light curves
\cite[]{Drave_2013,Rampy_2009,Bozzo_2017}. About half of SFXTs show
X$-$ray pulsations  \citep{Sidoli_2013}, proving that many (if not
most) contain neutron stars.  It is believed that some SFXTs are
descendants of BeHMXBs \cite[]{Liu_2011}. For instance, systems that
were classified as SFXTs in X-ray showed periodic flares in their
X-ray folded light curves \cite[]{Jain_2009}, and strong H$\alpha$
emission in optical as in Be systems (see Figure 1 in
\citealt{Hare_2019} and Figure 10 in \citealt{Coe_2021}).  However,
other SFXTs were noticed to house SG stars and accrete via Bondi
accretion \cite[]{Shakura_2014}.  

The optical and IR emission of HMXBs provides important insight into the nature of the stellar companion, as well as the warm gas in these systems which likely comprise the bulk of the material ejected by the star.  However, given the high $N_{\rm H}$ observed towards most Galactic HMXBs, emission in these wavebands is often hard to detect for such sources.  Therefore, studying extragalactic HMXBs is important for understanding the origin of the optical and 
IR emission of such objects.  IC 10 X$-$2 is a HMXB first identified
due to a massive X-ray outburst ($\sim100\times$ increase in flux) it
produced 2003, during which it was the second X-ray brightest object
in the IC 10 galaxy \cite[]{wang_2005,kwan_2018}. \cite{Laycock_2014}
used the Chandra position (RA: 00:20:20.94, Dec: 59:17:59.0) to find
the optical counterpart, analyzed the X-ray and optical spectra of IC
10 X$-$2, and concluded that the companion in the system is a blue
SG. The optical spectrum showed a smooth continuum and revealed strong
H$\alpha$ emission and a complex of Fe{\sc ii} permitted and forbidden
emission lines, similar to B[e] stars, i.e., B$-$type stars with
forbidden emission lines. The He{\sc i} emission and forbidden lines
IC 10 X$-$2 studied by \cite{Laycock_2014} have also been detected in
some sgB[e] systems \cite[]{Lamers_1998}. 

The hard X-ray spectrum of IC 10 X$-$2 suggests that the compact
object is a NS. A more recent study by \cite{kwan_2018} of the optical
and IR light curve and spectrum of IC 10 X$-$2 revealed that the
properties of the He{\sc i}, Paschen$-\gamma$, and Paschen$-\beta$
emission lines, and the mid$-$IR colors and magnitudes, showed strong
similarities to those in Luminous Blue Variables (LBV) or LBV
candidates (LBVc), and hence suggested the companion of IC 10 X$-$2 is
an LBV/LBVc.  

In Section \ref{sec:data}, we present the Zwicky Transient Facility
(ZTF) and Las Cumbers Observatory (LCO) data analyzed in this work. We
present in Section \ref{sec:Data analysis} the light curves and
color-magnitude diagrams of this object, which we use to identify
flares (\S\ref{sec:flare}) and periodic variations in both the
brightness and color of this source (\S\ref{sec:periodicity}).  In
Section \ref{sec:interpretation}, we discuss the physical implications
of our findings, and summarize our results in \S\ref{sec:conc}).  

\begin{table*}[bht]
\caption{Description of the ZTF, LCO, Gaia DR3 filters.}
\begin{tabular}{lllllllll}\hline\hline
 & \multicolumn{2}{c}{ZTF \footnote{\citet{Lau_2021}}} &
  \multicolumn{3}{c}{LCO \footnote{\url{https://lco.global/observatory/}}}
  & \multicolumn{3}{c}{Gaia DR3} \footnote{\url{https://gea.esac.esa.int/archive/}} \\
  \hline
Cadence & \multicolumn{2}{c}{$\sim$ 1 day} & \multicolumn{3}{c}{$\sim$
  1 week} & &  &  \\ 
Filter & $z_g$ & $z_r$ & $g^\prime$ & $r^\prime$ & $i^\prime$ & bp & g
& rp \\
$\lambda_{\rm{central}}$ (\AA) & 4722 & 6340 & 4770 & 6215 & 7545 &
5341.29 & 6605.35 & 7849.14 \\
FWHM (\AA) & 1282 & 1515 & 1500 & 1390 & 1290 & 2530.79 & 4365.35 &
3082.81 \\
MJD range (day) & 58252--59264 & 58254--59264 &
\multicolumn{3}{c}{58084--59262} &  &  &  \\ 
Number   of data points & 323 & 483 & 76 & 79 & 74 &  &  & \\ \hline
\end{tabular}
\label{tab:obsdata}
\end{table*} 

\section{Observational Data}
\label{sec:data}

In this section, we describe the data analyzed in this work, obtained
from ZTF (Section \ref{sec:zwicky}), and the global telescope network
of Las Cumbres Observatory (LCO; Section \ref{sec:LCO}). A summary of
the data sets used in this work is provided in Table
\ref{tab:obsdata}.

\subsection{Zwicky Transient Facility} 
\label{sec:zwicky}

ZTF is an optical time-domain astronomical survey that uses an
enhanced CCD camera attached to the Samuel Oschin Telescope at the
Palomar Observatory in California, United States
\cite[]{Bellm_2019}. It uses three filters ($z_g$, $z_r$, and $z_i$)
with improved sensitivity compared to that in Sloan Digital Sky Survey
(SDSS), and has the ability to request observations a few hours apart
using these filters. The telescope scans the entire Northern sky at
night with a cadence of two days. Each data release reports selected
automatically detected targets' light curves; the MJD of a detection,
the corresponding magnitude and error, and the target's position in
the sky. We acquired three filters’ data, $z_g$, $z_r$, and $z_i$, of
IC 10 X$-$2 from the Public Data Release 6 \cite[]{web_ztf_DR},
retrieved through NASA/IPAC Infrared Science Archive (IRSA)
\cite[retrieved on 23 October 2021]{web_IRSA}. Each filter data set is
labeled with an identification number (OID) as listed in Table
\ref{tab: ZTF data}. The angular distance search was less than
1$\arcsec$ from the target’s right ascension and declination. The
$z_i$ filter was not included in further analyses, as there were only
2 detections within our selected time frame (see Table
\ref{tab:obsdata}).

\begin{table*}[]
\centering
\caption{Properties of ZTF data on IC 10 X$-$2 obtained between 2018
  May 14 (MJD 58254) and 2019 October 15 (MJD 58771). OID refers to
  the identification number in the ZTF DR 6 catalog.} 
\begin{tabular}{lllll} \hline \hline 
OID & Filter & Number of observations & RA (deg) & Dec (deg) \\ \hline 
806103100029923 & $z_{g}$ & 278 & 5.086978 & 59.29979 \\
1809111300028700 & $z_{g}$ & 45 & 5.086909 & 59.29979 \\
806203100063307 & $z_{r}$ & 827 & 5.087091 & 59.29981 \\
1809211300074740 & $z_{r}$ & 84 & 5.087093 & 59.29979 \\
806303100062288 & $z_{i}$ & 2 & 5.087076 & 59.29979 \\ \hline 
\end{tabular}
\label{tab: ZTF data}
\end{table*}

\subsection{LCO}
\label{sec:LCO}

\begin{figure*}[] 
    \centering
    \subfigure{    
    \includegraphics[width=0.9\textwidth]{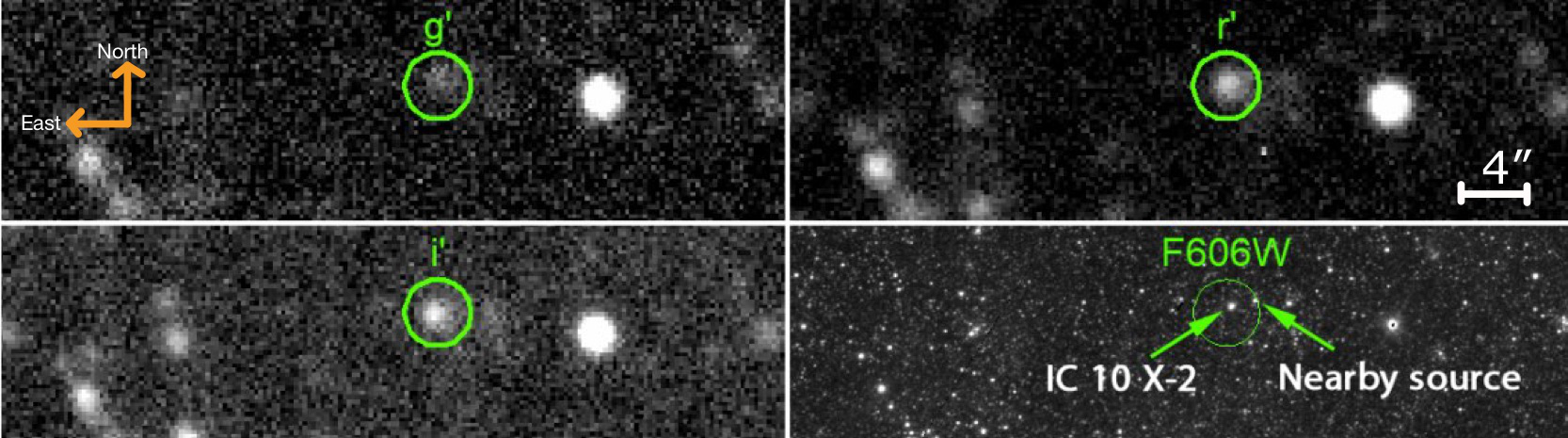}
    }  
    \caption{LCO images taken on MJD 59767 (2021 July 7), centered at
      IC 10 X$-$2's coordinates in $g^\prime$, $r^\prime$, $i^\prime$,
      and HST ACS/WFC $F606W$ filters. The $F606W$ image clearly shows
      IC 10 X$-$2 (brightest, on the left) and the nearby source
      (second brightest, on the right). The region radius in all
      images is 2$\arcsec$ and the filter used is indicated on each
      image.} 
    \label{fig: LCO images}
\end{figure*}

Observations of the source in three SDSS filters ($g^\prime$,
$r^\prime$ and $i^\prime$) were taken  $\sim$10 minutes apart between
2017 November 27 (MJD 58084) and 2021 July 31 (MJD 59426), using
5-minute exposures with the 1.0-meter telescope, and the 2.0-meter
Faulkes Telescope North equipped with the Muscat3 camera, of the Las
Cumbres Observatory (LCO) \cite[]{Brown_2013}. The 2.0-meter telescope
allows for simultaneous observations in $g^\prime$, $r^\prime$,
$i^\prime$, and $z^\prime$. The data were extracted and calibrated by
the X-ray Binary New Early Warning System (XB-NEWS) real-time data
analysis pipeline \cite[]{Russell_2019,Goodwin_2020}. XB-NEWS measured
IC 10 X-2 coordinates as RA:00:20:20.89 Dec:+59:17:59.2 with accuracy
$\lessapprox0.1\arcsec$. The statistical error associated with all LCO
magnitudes in this work is $<$0.09 mag.  

We show in Figure \ref{fig: LCO images} images of the IC 10 X-2 field
using LCO $g^\prime$, $r^\prime$, $i^\prime$, and a Hubble Space
Telescope Advanced Camera for Surveys/Wide Field Channel (HST ACS/WFC)
higher resolution image in the $F606W$ filter. We found two bright
sources within $2\arcsec$ of IC 10 X$-$2's coordinates
($\sim1.6\arcsec$ away) in the HST image. This suggests that
extraction of LCO photometry with a 2$\arcsec$ radius at IC 10 X$-$2's
coordinates should include a brightness contribution from the nearby
source. We searched in Gaia DR 3 \citep{Gaia_DR3} for the magnitude of
both sources (IC 10 X-2 and the nearby source) to quantify the nearby
source's contribution to the measured magnitude. The Gaia-$g$
magnitudes were reported in Gaia DR 3 as 19.00 mag for IC 10 X$-$2 and
20.54 mag for the nearby source, respectively, i.e., one-fourth of IC
10 X$-$2's flux. The average magnitude and color measured for the
brighter Gaia source is consistent with the average magnitude and
color measured in our LCO observations. Therefore, we conclude that
the brighter of the two Gaia sources dominate the total emission
measured in the observations we analyze which we associate with IC 10
X-2.

\begin{table*}[]
\centering
\caption{Gaia DR3 details of sources within 2$\arcsec$ of IC 10 X-2
  coordinates.}
\begin{tabular}{lllll} \hline\hline
Gaia ID & RA & Dec & Gaia $g$ (mag) & Gaia $bp-rp$ (mag) \\\hline
428242947042534784 & 5.086406836 & 59.29988915 & 20.54 & NA \\
428242947052289664 & 5.087189803 & 59.29978945 & 19.00 & 1.67 \\\hline
\end{tabular}
\end{table*}

\section{Data Analysis and Results}
\label{sec:Data analysis} 

In this Section, we first discuss the observed IC 10 X$-$2 multi-band
light curve in Section \ref{subsec: LC} and color in Section
\ref{sec:color}, then we identify flares in Section \ref{sec:flare},
and lastly, we search for periodicity in the light curves in Section
\ref{sec:periodicity}.  

\subsection{Multi-Band Light Curve}
\label{subsec: LC}

\begin{figure}[h] 
    \centering\includegraphics[width=0.49\textwidth]{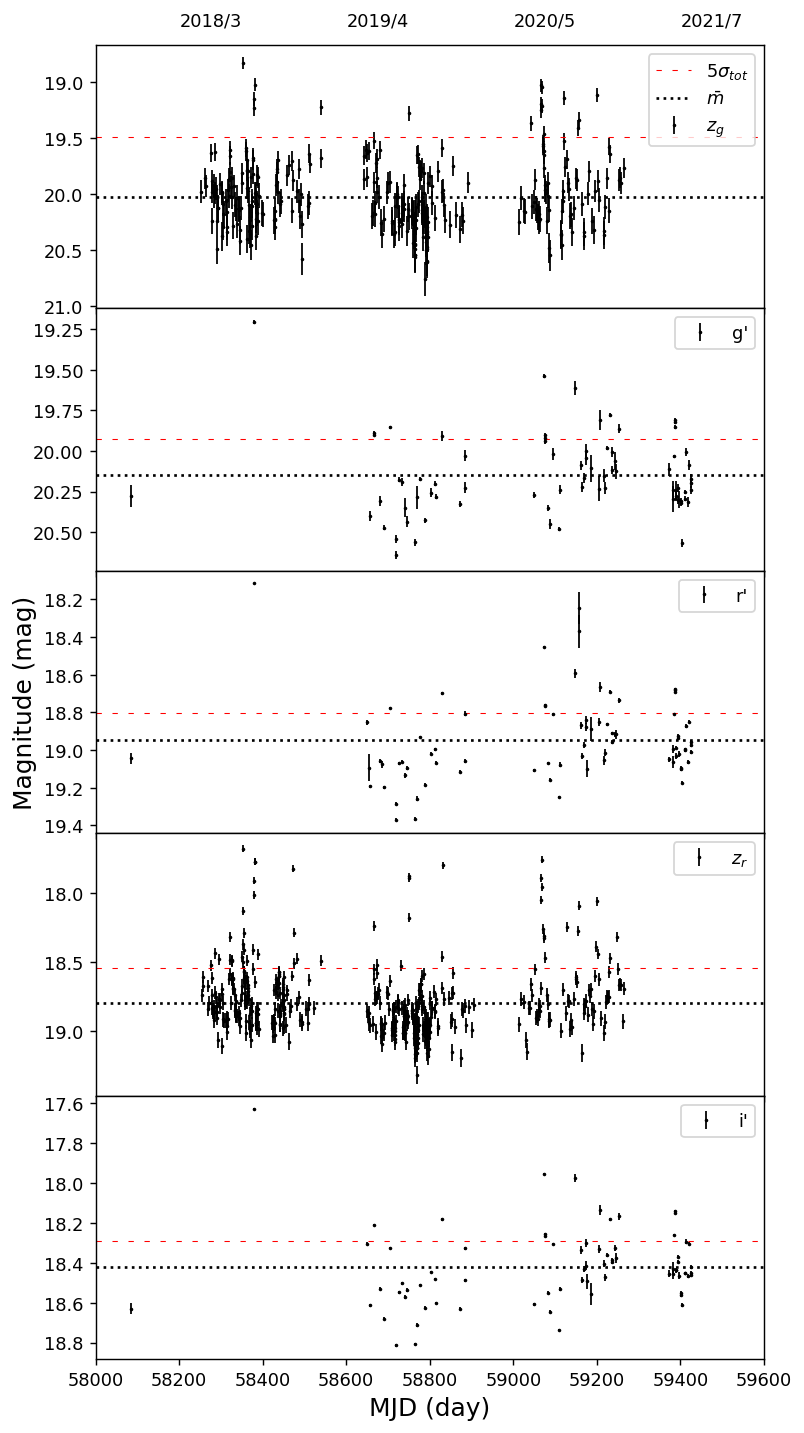}
    \caption{Apparement magnitude of IC 10 X$-$2 in $g^\prime$, $z_g$,
      $r^\prime$, $z_r$, and $i^\prime$ filters during the range of
      MJDs specified in Table \ref{tab:obsdata}. The dashed and dotted
      lines reflect the flare threshold (5$\sigma$) and average
      apparent magnitude $\bar{m}$, respectively, as calculated in
      \S\ref{sec:flare}.}
    \label{fig: Light Curve}
\end{figure}

Figure \ref{fig: Light Curve} shows the apparent $i^{\prime}$,
$r^{\prime}$ / $z_r$, and $g^\prime$ / $z_g$ magnitudes of IC 10 X$-$2
as measured on different epochs in the observations described in
\S\ref{sec:data}.  In all five filters, there were epochs when the
source was brighter (lower apparent magnitude) than average, with the
observed variations ($\sim1-1.5$ magnitudes in the $z_g$ /
$g^{\prime}$, $z_r$ / $r^{\prime}$ filters, and $\sim$1 in $i^\prime$)
much larger than the typical errors in these measurements ($\sim$0.08
for $z_g$ / $z_r$, $\sim$0.02 for $g^{\prime}$ / $r^{\prime}$, and
$\sim$0.05 for $i^{\prime}$) -- suggesting these changes are intrinsic
to the source.

\subsection{Observed Color}
\label{sec:color} 

\begin{figure}[tbh]
    \begin{center}
    \includegraphics[width=0.475\textwidth]{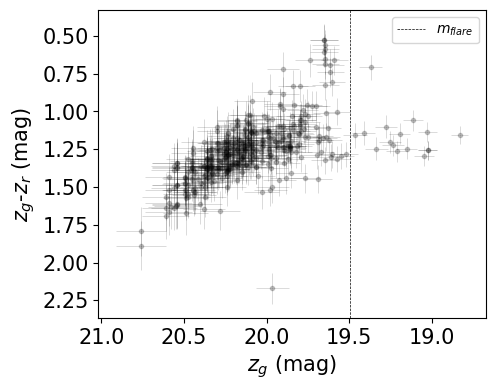}   
    \includegraphics[width=0.475\textwidth]{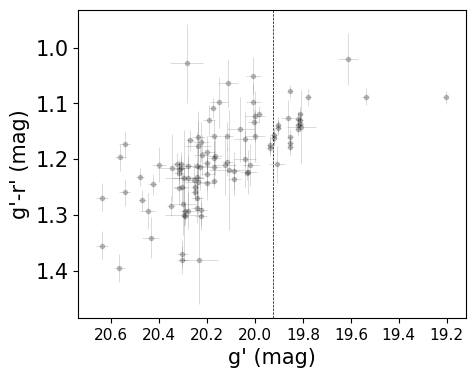}
    \end{center}
    \vspace*{-0.5cm}
    \caption{Observed color magnitude diagrams $z_{g}-z_{r}$
      vs. $z_{g}$ ({\it top}), and $g^\prime-r^\prime$ vs. $g^\prime$
      ({\it bottom}) of IC 10 X-2. $m_{\rm flare}$ refers to the flare
      magnitude threshold calculated in \S\ref{sec:flare}.}
    \label{fig:CMDs} 
\end{figure}

Figure \ref{fig:CMDs} shows the apparent color ($z_{g}-z_{r}$,
$g^\prime-r^\prime$) as a function of the apparent magnitude ($z_g$
and $g^\prime$, respectively) measured on a particular epoch. As shown
in this Figure, it appears that when IC 10 X-2 is brighter (lower
absolute magnitude), its emission is also bluer (lower
color). Furthermore, it also appears that above a certain brightness,
the color of IC 10 X-2 is roughly constant.

To determine the statistical significance of the relationship between
color ($z_g-z_r$) and apparent magnitude $z_g$ described above, we
calculated their weighted correlation coefficient (WCC):
\begin{eqnarray}
    \label{eqn:wcc}
    {\rm WCC} & = & \frac{C_{w}(m,\Delta
      m,\sigma_{m})}{\sqrt{C_{w}(m,m,\sigma_{m}) \, C_{w}(\Delta m,
        \Delta m, \sigma_{m})}}, 
\end{eqnarray}
where $m$ is the apparent magnitude, $\Delta m$ is the apparent
color. $M_w$ is the weighted mean: 
\begin{eqnarray}
\label{eqn:mw}
 M_{w} (m, \sigma_{m}) & = & \frac{\sum m \,\sigma_{m}}{\sum \sigma_{m}},
\end{eqnarray}
and $C_{w}$ is the weighted covariance:
\begin{eqnarray}
 C_{w} (m, \Delta m, M_w) & = & \frac{\sum \sigma_{m} (m-M_{w})
   (\Delta m - M_{w})}{\sum \sigma_{m}}. 
\end{eqnarray}
For $z_g-z_r$, we find that when IC 10 X-2 is dimmer (apparent
magnitude $z_g \geq 19.5$, the threshold for ``flares'' discussed in
\S\ref{sec:flare}, WCC=0.73 -- suggestive of a statistically
significant correlation between color ($z_g-z_r$) and apparent $z_g$
magnitude, while when IC 10 X-2 is brighter ($z_g < 19.5$), this
correlation is significantly weaker or not present.  These results are
consistent with the observation discussed above, possible physical
origins of this connection is discussed in \S
\ref{sec:interpretation}.

\subsection{Flare Identification and Properties}
\label{sec:flare}
As mentioned in \S\ref{subsec: LC}, we frequently detected variations
in the apparent magnitude of IC 10 X-2 significantly larger than the
statistical error of individual measurements.  Similar behavior in the
optical and infrared emission of this source was previously reported
by \citet{kwan_2018}, who associated those epochs with an apparent
magnitude 5$\sigma$ smaller than average with ``flares'' from the
source. We use a similar method to determine the threshold for flares
for a particular filter in the observations analyzed here (Table
\ref{tab:obsdata}), which we do by:
\begin{enumerate}
    \item fitting the light curve in a particular filter with a
      constant $\bar{m}$, which has an associated (statistical)
      uncertainty $\sigma_{\rm stat}$,
    \item determine the (systematic) uncertainty in $\bar{m}$ by
      calculating the average error in apparent magnitude,
      $\sigma_{\rm sys}$, for all epochs where $m$ is within
      $\bar{m}\pm3\sigma_{\rm stat}$,
    \item calculate the total uncertainty in $\bar{m}$ as:
      \begin{eqnarray}
        \label{eqn:error}
        \sigma_{\rm tot} & = &\sqrt{ \sigma_{\rm stat}^2 + \sigma_{\rm sys}^2},
    \end{eqnarray}
    \item and, finally, set the flare magnitude threshold to be:
    \begin{eqnarray}
        \label{eqn:mflare}
        m_{\rm flare} & \equiv & \bar{m}-5\sigma_{\rm tot}.
    \end{eqnarray}
\end{enumerate}
The resultant values of these quantities for each of the observed
filters are given in Table \ref{tab:flare_threshold}, and both
$\bar{m}$ and $m_{\rm flare}$ are indicated in the light curves shown
in Figure \ref{fig: Light Curve}.

\begin{table}[tb]
\caption{Quantities used to calculate the magnitude threshold for IC
  10 X-2 in the observed filters}
\begin{center}
\begin{tabular}{llllll} 
\hline
\hline
Filter & $\bar{m}$ & $\sigma_{\rm stat}$ & $\sigma_{sys}$ &
$\sigma_{tot}$ & $m_{flare}$ \\ 
$\cdots$ & mag & mag & mag & mag & mag \\
\hline
$z_g$ & 20.02 & 0.02 & 0.10 & 0.11 & 19.49 \\
$g^\prime$ & 20.15 & 0.03 & 0.03 & 0.04 & 19.93 \\
$r^\prime$ & 18.95 & 0.03 & 0.01 & 0.03 & 18.80 \\
$z_r$ & 18.79 & 0.01 & 0.05 & 0.05 & 18.54 \\
$i^\prime$ & 18.42 & 0.02 & 0.01 & 0.03 & 18.29 \\
\hline
\end{tabular}
\end{center}
\label{tab:flare_threshold} 
\end{table}

\begin{figure}[tb]
    \centering
    \includegraphics[width=0.45\textwidth]{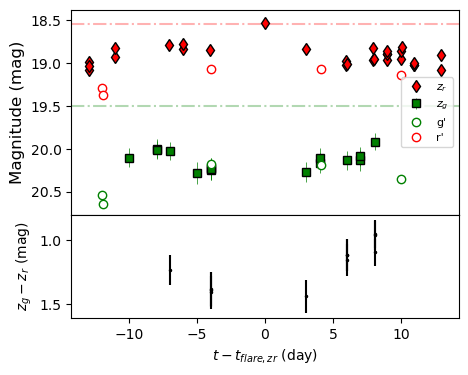}
    \caption{Multi-band light curve ({\it top}) and color ({\it
        bottom}) of IC 10 X-2 during W5 (Table \ref{tab:flare_fit}).} 
    \label{fig:w5_lc}
\end{figure}

\begin{figure*}[tbh]
    \centering    
    \subfigure[W1]{\includegraphics[width=0.45\textwidth]{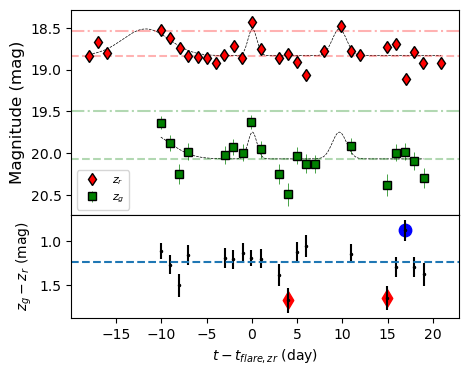}
    }     
    \subfigure[W2]{\includegraphics[width=0.45\textwidth]{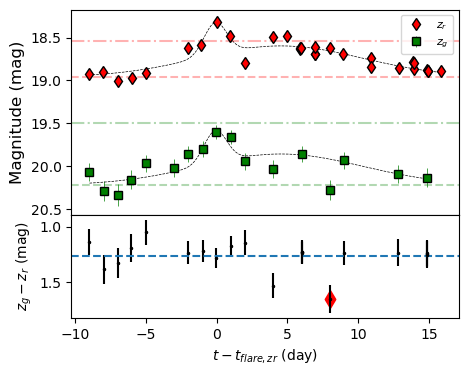}  
    } 
    \subfigure[W3]{\includegraphics[width=0.45\textwidth]{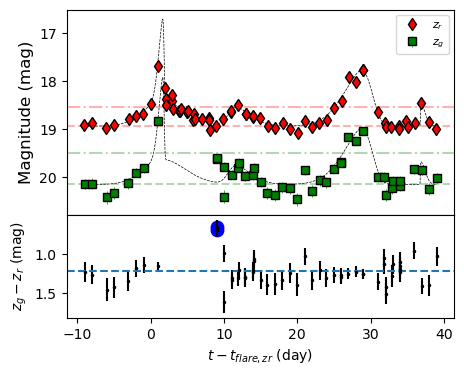} 
    }     
    \subfigure[W4]{    
    \includegraphics[width=0.45\textwidth]{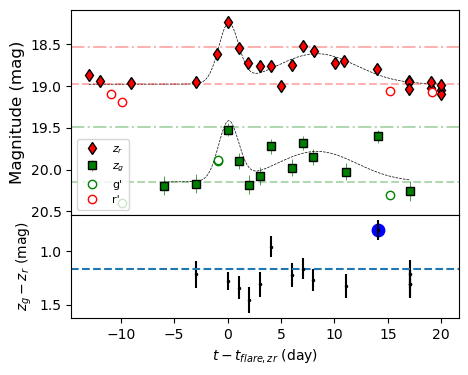}
    }     
    \subfigure[W6]{    
    \includegraphics[width=0.45\textwidth]{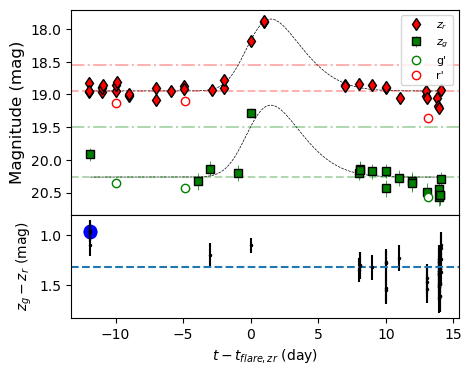} 
    }        
    \subfigure[W7]{     
    \includegraphics[width=0.45\textwidth]{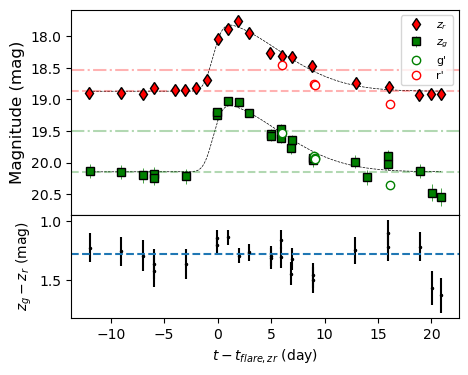}  
    }           
    \caption{Apparent magnitude ({\it top} panel) and color ({\it
        bottom} panel) of IC 10 X-2 during each window of flaring
      activity described in \S\ref{sec:flare}. In the top panels, the
      solid lines indicate the skew-normal fits to the light curve
      (Table \ref{tab:flare_fit}), the dashed likes the quiescent
      magnitude $m_q$ (Table \ref{tab:flare_fit}) and the flare
      threshold $m_{\rm flare}$ (Table \ref{tab:flare_threshold}) for
      each filter.  In the bottom panels, the dashed line indicates
      the quiescent color $m_{q,zg}-m_{q,zr}$, while epochs with color
      less / greater than $3\sigma$ of the quiescent color level are
      marked with blue circles / red diamonds, respectively.}
    \label{fig:flares}
\end{figure*}

As discussed in \S\ref{sec:periodicity}, the emission from IC 10 X-2
varies periodically with a period $P_{\rm orb} \sim 26.5~{\rm d}$.  To
identify when IC 10 X-2 exhibited flaring behavior, we defined a
temporal window of width $P_{\rm orb}$ around each epoch with $m \leq
m_{\rm flare}$.  We then merged together overlapping windows --
resulting in several distinct episodes of flaring behavior listed in
Table \ref{tab:flare_fit}.  As shown in Fig. \ref{fig:w5_lc}, the
detection of a flare during W5 was quite marginal, in contrast to the
numerous strong flares during the other windows of activity as shown
in Figure \ref{fig:flares}.

To measure the properties of these flaring episodes, we modeled the
$z_r$ light curve during each window (but W5, due to a lack of data)
as a sum of skew-normal functions\footnote{whose implementation within
  {\tt scipy} is described at
  \url{https://docs.scipy.org/doc/scipy/reference/generated/scipy.stats.skewnorm.html}
  plus a constant:
\begin{equation}
    m(x)=\sum_{i} 2A_i\phi\left(\frac{x-\Delta t_i}{\sigma_i}\right)
    \Phi \left(a_i \frac{x-\Delta t_i}{\sigma_i} \right) + m_q 
    \label{eqn:skew-normal}
\end{equation}
where:
\begin{itemize}
    \item $x$ is the time elapsed from the first $z_r$ magnitude
      measurement in the window, 
    \item $\phi$ is the Gaussian probability density function, and 
    \item  $\Phi$ is the cumulative distribution function of a Gaussian.
\end{itemize}
We used the non-linear least squares method, as implemented by the
{\tt curve\_fit} function in {\sc SciPy}, to determine the value of
the following quantities  
\begin{itemize}
    \item $A_i$, the amplitude of a particular skew-normal component,
    \item  $\Delta t_i$, the offset of a particular skew-normal
      component from the beginning of the window, 
    \item $\sigma_i$, the width of a particular skew-normal component,
    \item $a_i$ is the skewness parameter of a particular skew-normal
      componant, and  
    \item $m_q$, the quiescent $z_r$ magnitude of IC 10 X-2 during this window.
\end{itemize}
which minimized the $\chi^2$, estimateing the uncertainty of the
fitted value of each parameter ($\sigma$) using the diagonal elements
of the resultant covariance matrix.  For each Window, we added
skew-normal components until the fitted amplitude $A_i$ became
consistent within $3\sigma$ with zero ($A_i\leq 3\sigma_{A}$).
Furthermore, we initially allowed the skewness $a_i$ of each
skew-normal component to vary, but in those cases where $a_i$ was
consistent with zero ($a_i \leq \sigma_a$), we set $a_i \equiv 0$ in
order to minimize the number of model parameters.  As listed in Table
\ref{tab:flare_fit}, a wide range of amplitudes ($A_i$), durations
(width $\sigma_i$), and shape (skewness $a_i$) of the changes in
magnitude were observed during these flaring episodes -- with no
significant correlations observed between the values of these
different parameters.}  

\begin{table*}[tb]
\centering
\caption{Values of the ``best-fit'' model parameters to the observed
  apparent magnitude of IC 10 X-2 during falring "windows", number of
  data points $N$, and model $\chi^2$, in each window}
\begin{tabular}{cccccccccc}
\hline
\hline
Window & Filter & $t_{\rm flare,z_r}$ & $A$ & $\Delta t$ & $\sigma$ &
a & $m_{q}$  & N & $\chi^{2}$ \\ 
$\cdots$ & $\cdots$ & [MJD] & [mag] & [days] & [days] & $\cdots$ &
[mag] & $\cdots$ & $\cdots$ \\ 
\hline
\multirow{4}{*}{W1} & \multirow{3}{*}{$z_r$} &
\multirow{4}{*}{58286.47} & $0.80\pm0.20$ & $-11.73\pm0.49$ &
$2.59\pm0.42$ & $\equiv0$ & \multirow{3}{*}{$18.83\pm0.01$} &
\multirow{3}{*}{29} & \multirow{3}{*}{85.91} \\ 
 &  &  & $0.80\pm0.12$ & $0.06\pm0.25$ & $0.49\pm0.14$ & $\equiv0$ &  &  &  \\
 &  &  & $0.80\pm0.14$ & $9.68\pm0.30$ & $0.79\pm0.17$ & $\equiv0$ &  &  &  \\
 & $z_g$ &  & $\cdots$ & $\cdots$ & $\cdots$ & $\cdots$ &
$20.07\pm0.02$ & 20 & 40.10 \\ 
 \hline
\multirow{3}{*}{W2} & \multirow{2}{*}{$z_r$} &
\multirow{3}{*}{58321.47} & $0.93\pm0.13$ & $-0.1\pm0.11$ &
$0.80\pm0.13$ & $\equiv0$ & \multirow{2}{*}{$18.96\pm0.06$} &
\multirow{2}{*}{30} & \multirow{2}{*}{56.27} \\ 
 &  &  & $0.91\pm0.15$ & $4.52\pm0.55$ & $5.80\pm1.20$ & $\equiv0$ &  &  &  \\
 & $z_g$ &  & $\cdots$ & $\cdots$ & $\cdots$ & $\cdots$ &
$20.22\pm0.02$ & 17 & 23.60 \\ 
 \hline
\multirow{6}{*}{W3} & \multirow{5}{*}{$z_r$} &
\multirow{6}{*}{58352.33} & $2.28\pm0.76$ & $1.85\pm0.05$ &
$0.69\pm0.15$ & $-7.38\pm4.33$ & \multirow{5}{*}{$18.93\pm0.01$} &
\multirow{5}{*}{65} & \multirow{5}{*}{167.87} \\ 
 &  &  & $0.98\pm0.14$ & $-0.31\pm0.80$ & $3.82\pm0.66$ &
$1.34\pm0.86$ &  &  &  \\ 
 &  &  & $0.58\pm0.07$ & $10.58\pm0.19$ & $2.53\pm0.39$ &
$3.22\pm1.47$ &  &  &  \\ 
 &  &  & $1.72\pm0.06$ & $30.21\pm0.10$ & $3.26\pm0.16$ &
$-3.38\pm0.45$ &  &  &  \\ 
 &  &  & $0.60\pm0.11$ & $36.78\pm7.31$ & $0.62\pm4.01$ &
\multicolumn{1}{l}{$18.17\pm0.01$} &  &  &  \\ 
 & $z_g$ &  & $\cdots$ & $\cdots$ & $\cdots$ & $\cdots$ &
$20.15\pm0.01$ & 45 & 173.45 \\ 
 \hline
\multirow{3}{*}{W4} & \multirow{2}{*}{$z_r$} &
\multirow{3}{*}{58666.43} & $1.75\pm0.09$ & $0.09\pm0.07$ &
$0.93\pm0.07$ & $\equiv0$ & \multirow{2}{*}{$18.98\pm0.02$} &
\multirow{2}{*}{25} & \multirow{2}{*}{59.68} \\ 
 &  &  & $0.91\pm0.07$ & $8.35\pm0.36$ & $3.74\pm0.48$ & $\equiv0$ &  &  &  \\
 & $z_g$ &  & $\cdots$ & $\cdots$ & $\cdots$ & - & $20.15\pm0.03$ & 13
& 54.43\\ 
 \hline
\multirow{2}{*}{W5} & $z_r$ & \multirow{2}{*}{58730.4} & $\cdots$ &
$\cdots$ & $\cdots$ & $\cdots$ & $\cdots$ & $\cdots$ & $\cdots$ \\ 
 & $z_g$ &  & $\cdots$ & $\cdots$ & $\cdots$ & $\cdots$ & $\cdots$ &
$\cdots$ & $\cdots$ \\ 
 \hline
\multirow{2}{*}{W6} & $z_r$ & \multirow{2}{*}{58751.33} &
$1.79\pm0.17$ & $-0.15\pm0.22$ & $3.18\pm0.26$ & $2.30\pm0.56$ &
$18.95\pm0.01$ & 38 & 85.48 \\ 
 & $z_g$ &  & $\cdots$ & $\cdots$ & $\cdots$ & $\cdots$ &
$20.26\pm0.03$ & 20 & 50.30 \\ 
 \hline
\multirow{2}{*}{W7} & $z_r$ & \multirow{2}{*}{59068.38} &
$1.38\pm0.03$ & $-0.44\pm0.06$ & $5.97\pm0.22$ & $8.41\pm1.08$ &
$18.87\pm0.02$ & 21 & 46.60 \\ 
 & $z_g$ &  & $\cdots$ & $\cdots$ & $\cdots$ & $\cdots$ &
$20.15\pm0.02$ & 26 & 53.55 \\\hline 
\end{tabular}
\label{tab:flare_fit}
\end{table*}

As mentioned in \S\ref{sec:color}, the emission from IC 10 X-2 is
observed to have a roughly constant value of $z_g - z_r$ when IC 10
X-2 is brighter than the flare threshold calculated above
(Fig.\ \ref{fig:CMDs}).  To quantify how well our $z_r$ model
reproduces the $z_g$ emission during these Windows, we fit the $z_g$
light-curve with the same functional form (Equation
\ref{eqn:skew-normal}) and parameters (Table \ref{tab:flare_fit}
derived from the $z_r$ data, save for the quiescent magnitude $m_q$
which was allowed to vary.  The resultant values of $m_q$ and $\chi^2$
for $z_g$ are reported in Table \ref{tab:flare_fit}.

As indicated in Table \ref{tab:flare_fit}, this model resulted in fits
with a reduced $\chi^2 \sim 1.5 - 4$.  As shown in Figure
\ref{fig:flares}, the largest deviations from the model are typically
isolated data points -- with a significant change in magnitude, often
only in one filter ($z_g$ or $z_r$), detected in $\sim1-2$ epochs.
These chromatic variations in the light curve could be indicative of
variability on timescales shorter than the $\sim 0.5-1$~days
separating observations in these filters.  Such chromatic variations
would result in significant changes in the color ($z_g-z_r$), and
epochs whose color is $\gtrsim3\sigma$ bluer / redder than quiescence
during each window are indicated as blue / red points in the color
panels of Figure \ref{fig:flares}.  The possible nature of these
variations will be discussed in \S\ref{sec:interpretation}.

\subsection{Periodic Variability}
\label{sec:periodicity}
 
\begin{figure*}[tbh]
\begin{center}
\includegraphics[width=0.475\textwidth]{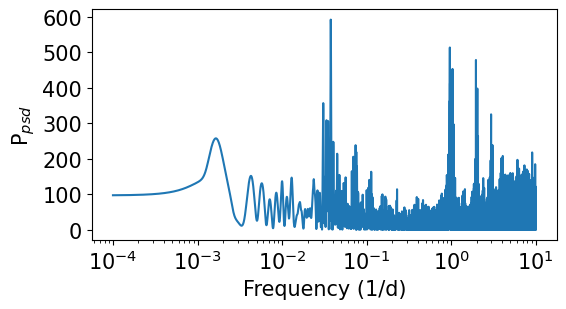}
\includegraphics[width=0.475\textwidth]{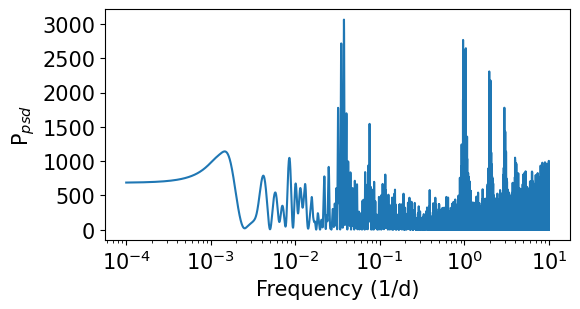} 
\includegraphics[width=0.475\textwidth]{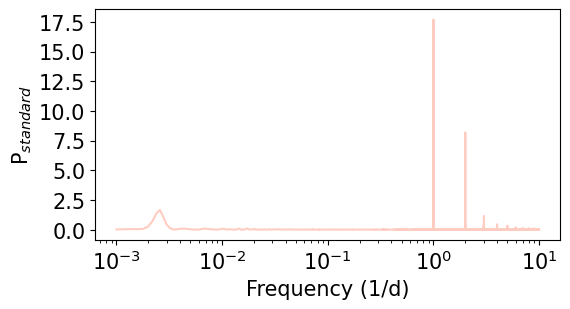}
\includegraphics[width=0.475\textwidth]{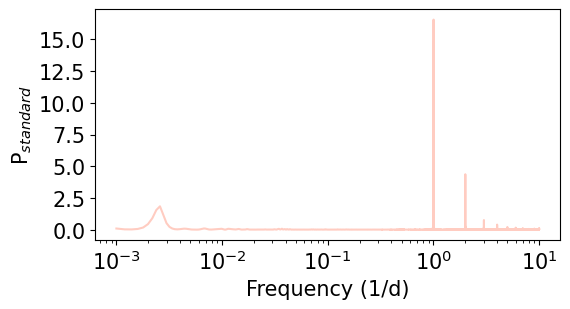}
\end{center}
\caption{Lomb-Scargle periodograms of the measured $z_g$ ({\it left})
  and $z_r$ ({\it right}) apparent magnitudes of this source ({\it top
    row}) and of the ``window'' functions as defined in
  \S\ref{sec:periodicity} {\it bottom row}).}
\label{fig:periodograms}
\end{figure*}

To determine if there are periodic variations in the emission from IC
10 X-2, we calculated the Lomb-Scargle periodogram
\citep{Lomb_1976,Scargle_1982} of its observed $z_g$ and $z_r$ light
curves.  A commonly used technique for identifying periodic signals in
light curves derived from unevenly spaced observations, in this method
the apparent magnitudes are essentially fit with a model (e.g.,
Equation 35 in \citealt{VanderPlas_2018}):
\begin{eqnarray}
\label{eqn:lstest}
m(t;f) = A_f \sin(2\pi f(t-\phi_f))
\end{eqnarray}
where $f$ is the trial frequency of the periodic variations $(f \equiv
1/P$ where $P$ is the period$)$, where $A_f$ and $\phi_f$ minimize the
$\chi^2(f)$ of the fit at frequency $f$.  The power ${\mathcal P}_{\rm
  psd}$ of a periodic signal with frequency $f$ in the data is defined
to be (e.g., Equation 37 in \citet{VanderPlas_2018}):
\begin{eqnarray}
    \label{eqn:psd}
    {\mathcal P}_{\rm psd}(f) & = & \frac{1}{2}(\chi^2_{\rm ref} - \chi^2(f)),
\end{eqnarray}
where $\chi^2_{\rm ref}$ is the $\chi^2$ resulting from performing
such a fit on the ``window,'' or reference, function defined to be a
constant magnitude at each observation epoch.  Using the {\tt
  LombScargle} class within {\sc astropy}, we searched for periodic
modulations with a frequency $10^{-4} \leq f \leq 10$~day$^{-1}$
($0.1~{\rm days} \leq P \leq 10^4~{\rm days}$), using a samples per
peak (number of trial frequency $f$ across the Nyquist frequency
derived from average spacing between observations) of 100.  The
resulting periodograms derived using all $z_g$ and $z_r$ observations
are shown in Figure \ref{fig:periodograms}.  Since frequency $f$ is
the only free parameter when calculating the periodogram, we define
the 95\% confidence interval on $f_{\rm max}$ ($P_{\rm max}$) as the
lowest and highest frequencies (periods) with power $\mathcal{P} >
\mathcal{P}_{\rm max}-1$ ($\Delta \chi^2 = 2$ relative to
$\mathcal{P}_{\rm max}$; Equation \ref{eqn:psd}).  The highest power
${\mathcal P}_{\rm psd}$, corresponding frequency $f_{\rm max}$ and
period $P_{\rm max}$), and the uncertainty in these latter two
quantities, are listed in Table \ref{tab:periodicity}.

\begin{table}[tbh]
\caption{Properties of the periodic signal identified from the
  Lomb-Scargle analysis of the apparent magnitude of IC 10 X-2.  All
  quantities are defined in \S\ref{sec:periodicity}}
\vspace*{-0.5cm}
\hspace*{-2.2cm}
\begin{tabular}{cccccc} \\ 
\hline
\hline
Filter & Flares? & $\mathcal{P}_{\rm FAL,99}$ & $\mathcal{P}_{\rm
  max}$ & $f_{\rm max}$ & $P_{\rm max}$ \\ 
 & & & & $10^{-2}{\rm day}^{-1}$ & days \\
\hline
$z_g$ & Y & 318.8 & 592.7 & $3.768_{-0.003}^{+0.002}$ &
$26.54_{-0.01}^{+0.02}$ \\ 
$z_r$ & Y & 1167.4 & 3063.1 & $3.762_{-0.001}^{+0.001}$ &
$26.591_{-0.007}^{+0.007}$ \\ 
\hline
$z_g$ & N & 78.24 & 77.65 & $3.755_{-0.007}^{+0.008}$ &
$26.63_{-0.06}^{+0.05}$ \\ 
$z_r$ & N & 202.3 & 540.9 & $3.769_{-0.003}^{+0.004}$ &
$26.53_{-0.03}^{+0.02}$ \\ 
\hline
\hline
\end{tabular}
\label{tab:periodicity}
\end{table}

As shown in Table \ref{tab:periodicity}, for both filters the maximum
power $\mathcal{P}_{\rm max}$ in the periodogram exceeds the 99\%
confidence false alarm level $\mathcal{P}_{\rm FAL, 99}$, which for a
periodogram with $N$ trial frequencies, is the power which
$\mathcal{P}$ whose probability of occurring by chance is $1/(N\times
100)$, and the frequencies $f_{\rm max}$ (period $P_{\rm max}$)
corresponding to $\mathcal{P}_{\rm max}$ agree to
$\lesssim10^{-3}$. Both $\mathcal{P}_{\rm max} > \mathcal{P}_{\rm
  FAL,99}$ and consistent values of $f_{\rm max}$ ($P_{\rm max}$)
strongly suggests the emission from IC 10 X-2 varies with a period of
$P \sim 26.5~{\rm days}$. 

However, it is possible that this seeming periodicity is an artifact
of the location of the IC 10 X-2 in the sky, and / or the cadence of
these observations.  To test the first scenario, we computed the
periodogram of five ZTF $z_r$ sources and four $z_g$ sources
$\lesssim30^{\prime \prime}$, and with a similar number of
observations and apparent magnitude, as IC 10 X-2.  None of these
sources show a statistically significant periodicity, let alone at
$P_{\rm max} \sim 26.5$ days, arguing against the first possibility
listed above.

To test the second possibility, we calculated the periodogram of the
``window'' function of the $z_g$ and $z_r$ observations of IC 10 X-2:
a time series with the same MJD as the actual observations of IC 10
X-2, but each epoch has the same apparent magnitude.  Since this is
the reference function used in Equation \ref{eqn:psd}, for this
periodogram we computed the ``standard'' normalization of the
power\footnote{As defined by the {\tt LombScargle} class in {\sc
    astropy}}:
\begin{eqnarray}
    \label{eqn:pstandard}
    \mathcal{P}_{\rm standard}(F) & \equiv & \frac{\chi^2_{\rm
        ref}-\chi^2(f)}{\chi^2_{\rm ref}} = \frac{2}{\chi^2_{\rm
        ref}}\mathcal{P}_{\rm psd}(F), 
\end{eqnarray}
to more easily identify frequencies $f$ where a periodicity may appear
to occur simply due to the cadence, or spacing, of the observation.
As shown in Figure \ref{fig:periodograms}, there are frequencies $f$
with significant power which are listed in Table \ref{tab:fp}. In the
``window" function, the frequencies with nearly (integer) frequencies
$f$ have significant power, and for $f \approx 1, 2, 3$ significant
power is also observed in the periodogram of IC 10 X-2.  However, the
window function contains no evidence for a signal at $f \approx f_{\rm
  max}$ present in the periodogram of IC 10 X-2 (Figure
\ref{fig:periodograms}, Table \ref{tab:fp}).  We further tested the
possibility of the observed $f_{\rm max}$ being an artifact of the
cadence of ZTF observations by randomizing the $z_g$ and $z_r$ light
curves $10^4$ times and calculating the resultant periodogram -- with
none of these randomized light curves having a $\mathcal{P}_{\rm max}$
comparable to that observed in the actual light curve of IC 10 X-2.
Therefore, based on these findings, we conclude that the observed
$\sim26.5$ day periodicity in the emission of IC 10 X-2 is intrinsic
to the source and not an instrumental artifact of the ZTF
observations.

\begin{table}[tb]
\centering
\caption{Frequencies ($f~{\rm day}^{-1}$) with significant
    power in the periodograms (Figure \ref{fig:periodograms}).}
\begin{center}
\begin{tabular}{cccc}
\hline
\hline
\multicolumn{2}{c}{$z_g$} & \multicolumn{2}{c}{$z_r$} \\
IC 10 X-2 & Window & IC 10 X-2 & Window \\
\hline
 $\cdots$ & 6.0030 & $\cdots$ & 6.0079 \\ 
 $\cdots$ & 5.0026 & $\cdots$ & 5.0080 \\
 $\cdots$ & 3.9999 & $\cdots$ & 4.0055 \\
 2.9678 & 3.0053 & 2.9705 & 3.0053 \\
 1.9651 & 2.0028 & 1.9678 & 2.0028 \\ 
 0.9650 & 1.0027 & 0.9651 & 1.0027 \\
 0.0376 & $\cdots$ & 0.0376 & $\cdots$ \\
\hline
\hline
\end{tabular}
\end{center}
\label{tab:fp}
\end{table}

\begin{figure*}[tb]
    \begin{center}
    \includegraphics[width=0.475\textwidth]{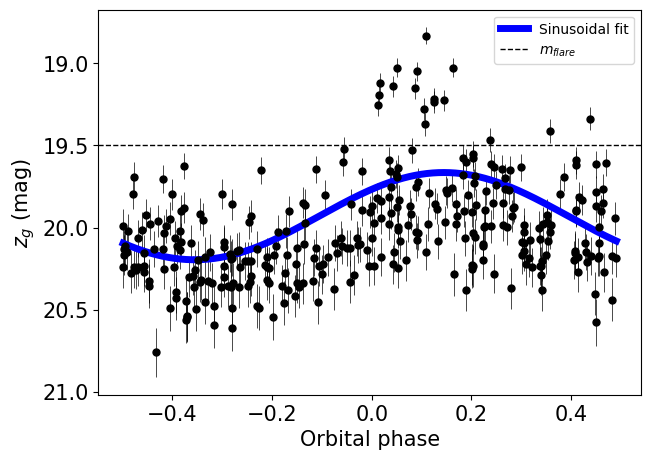}
    \includegraphics[width=0.475\textwidth]{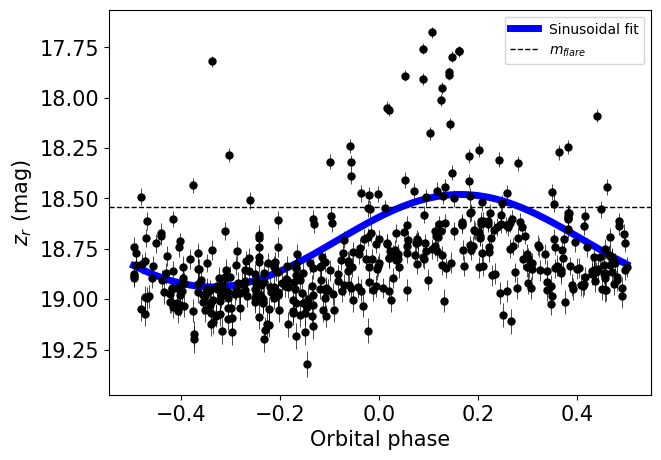}
    \end{center}
    \caption{ZTF $z_g$ ({\it left}) and $z_r$ ({\it right}) light
        curves of IC 10 X$-$2 folded with on a $P=26.54$ day period
        and phase $\phi({\rm MJD=58084})\equiv 0.$ In both figures,
        the dotted horizontal line indicates the flare threshold
        (Table \ref{tab:flare_threshold}), and the thick blue line
        indicates a sinusoidal fit to the apparent magnitude as a
        function of orbital phase $\phi$.}
    \label{fig:ztf_flc}
\end{figure*} 

\begin{figure*}[tbh]
    \begin{center}
    \includegraphics[width=0.32\textwidth]{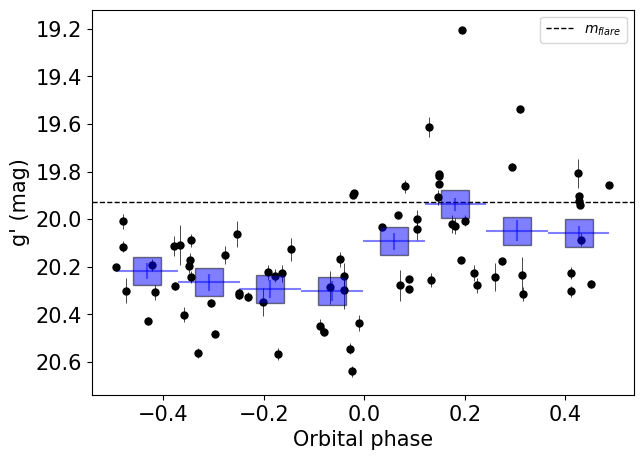}
    \includegraphics[width=0.32\textwidth]{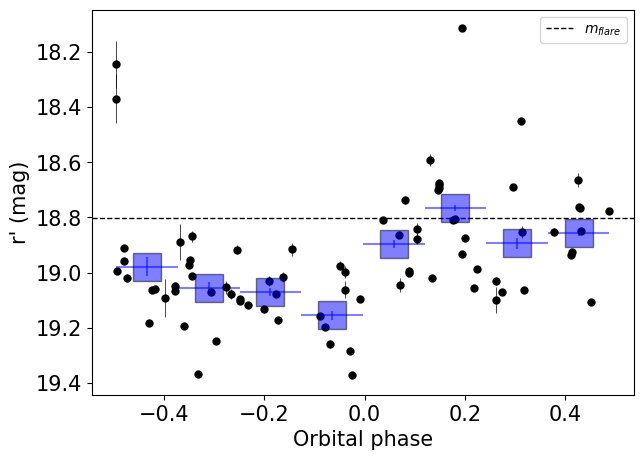}
    \includegraphics[width=0.32\textwidth]{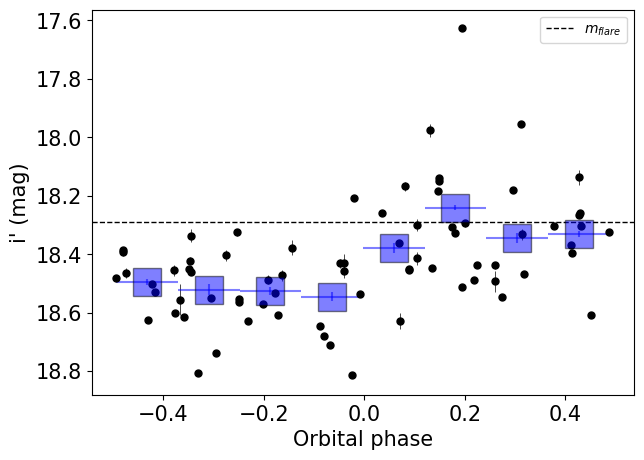}
    \end{center}
    \vspace*{-0.5cm}
    \caption{Light curves of $g^\prime$, $r^\prime$, and $i^\prime$
      with flares folded with a period of 26.54 days and the value
      $\phi_0$ derived from the folder $z_r$ light curve (Table
      \ref{tab:sin_parameters}).  The blue point indicate the binned
      folded light curve in these magnitudes.}
    \label{fig:lco_flc} 
\end{figure*}

To better study this periodicity, we folded the $z_g$ and $z_r$ light
curves with a period $P=26.54~{\rm days}$ ($P_{\rm max}$ for the $z_g$
data; Table \ref{tab:periodicity}) arbitrarily setting the phase at
MJD 58084 (the date of the first LCO observation) $\phi_0 \equiv 0$.
As shown in Figure \ref{fig:ztf_flc}, not only are measurements
brighter than the flare threshold ($m<m_{\rm flare}$) concentrated
within a fairly narrow range of phases ($z_g, \phi\sim0$ to $+0.2$;
$z_r, \phi \sim -0.1$ to $+0.2$), there appears to be a sinusoidal
variation in the apparent magnitude fainter than the flare threshold
($m > m_{\rm flare}$) in both filters.  We therefore fit the folded
apparent magnitudes $m$ with a sinusoidal function:
\begin{equation} 
\label{eqn:sin_fit}
m(\phi) = A \sin(2\pi (\phi-\phi_0)) + m_q
\end{equation}  
where $\phi$ is phase (-0.5 to 0.5), $A$ (mag) is the amplitude, and
$m_q$ (mag) is the midline magnitude.  The resultant parameters are
listed in Table \ref{tab:sin_parameters}.  Folding the $g^{\prime}$,
$r^{\prime}$, and $i^{\prime}$ light curves with the same 26.54 day
period and $\phi_0$ as found for $z_r$ (Figure \ref{fig:lco_flc})
finds a similar change in apparent magnitude with phase -- magnitudes
$m<m_{\rm flare}$ are concentrated at phases $\phi\sim0.2$, and IC 10
X-2 is fainter when $\phi < 0$ than $\phi > 0$. These results suggest
that both the flaring and quiescent emission from IC 10 X-2 are
periodic with the same $P\sim26.54~{\rm day}$ timescale.

\begin{table}[bt]
\caption{Model Parameters (Equation \ref{eqn:sin_fit}) derived from
  fitting the folded ZTF light curve of IC 10 X-2} 
\label{tab:sin_parameters}
\vspace*{-0.25cm}
\begin{center}
\begin{tabular}{ccc}
\hline
\hline
Parameter & $z_g$ & $z_r$ \\ 
A & $-0.265\pm0.008$ & $-0.230\pm0.003$ \\
$\phi_0$ & $0.105\pm0.005$ & $0.086\pm0.002$  \\
$m_q$ & $19.930\pm0.006$ & $18.710\pm0.002$ \\
$\chi^2$ / dof & 3122 / 320 & 17100 / 482 \\
 \hline
 \hline
\end{tabular}
\end{center}
\end{table}

\begin{figure*}
    \centering
    \includegraphics[width=0.475\linewidth]{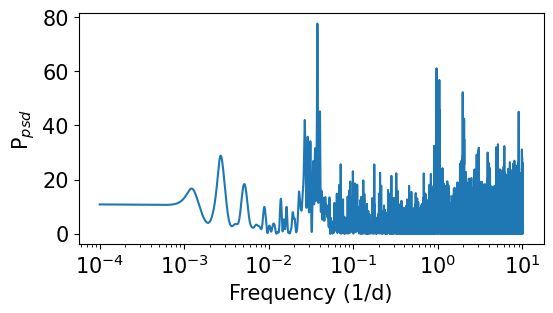}
    \includegraphics[width=0.475\linewidth]{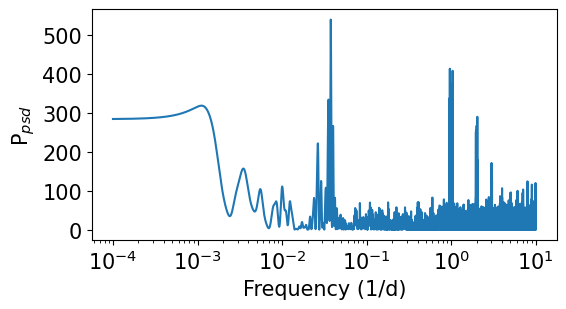}
    \caption{Lomb-Scargle periodograms of the measured $z_g$ ({\it
        left}) and $z_r$ ({\it right}) apparent magnitudes of IC 10
      X-2 on including epochs where $m > m_{\rm flare}$ (``quiescent''
      emission).} 
    \label{fig:periodogram_noflare}
\end{figure*}

While formally the sinusoidal model (Equation \ref{eqn:sin_fit}) does
not result in a good fit of the data, since the $\chi^2$ is much
greater than the degrees of freedom (dof; Table
\ref{tab:sin_parameters}), as shown in Figure \ref{fig:ztf_flc} this
model appears to reproduce changes in the apparent magnitude of IC 10
X-2 when $m>m_{\rm flare}$.  To determine if the emission from IC 10
X-2 is periodic in its ``quiescent'' state, we calculated the
periodogram of this source only using epochs where $m>m_{\rm flare}$.
As shown in Figure \ref{fig:periodogram_noflare}, the periodogram of
the quiescent emission of IC 10 X-2 shows significant power
$(\mathcal{P}_{\rm psd} \gtrsim \mathcal{P}_{\rm FAL,99})$ at
frequencies $f_{\rm max}$ (period $P_{\rm max}$) $<1\%$ defined than
observed in the full lightcurve of this source (Table \ref{tab:fp}.
Therefore, we conclude that both the ``quiescent'' and flare emission
from IC 10 X-2 periodically vary on a $P\sim 26.54~{\rm d}$ interval.

\begin{figure}[tbh]
    \includegraphics[width=0.4\textwidth]{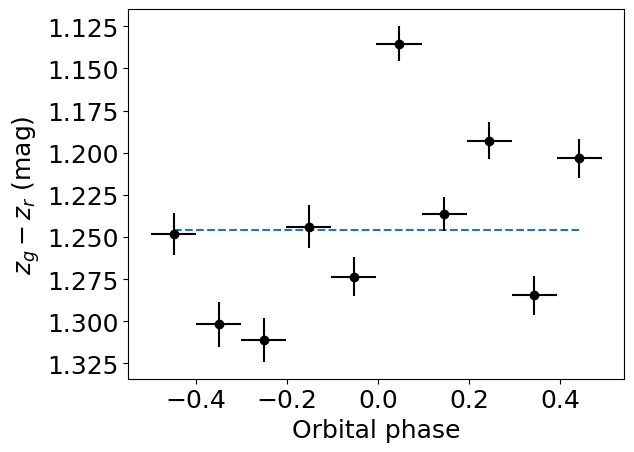}
    \caption{Average bserved $z_{g}-z_{r}$ color of IC 10 X-2 for
      phase bins for a 26.54 d period. The dashed horizontal line
      indicates the best-fit constant color of this source.}
    \label{fig:folded_color} 
\end{figure}

Finally, as shown in Table \ref{tab:sin_parameters}, the periodic
oscillations in the $z_g$ and $z_r$ emission of IC 10 X-2 have a
different amplitude $A$ and phase $\phi_0$.  As a result, we
calculated the ``quiescent'' $z_g - z_r$ color of IC 10 X-2 as a
function of phase $\phi$ assuming a 26.54~d period as above.  The
results are shown in Figure \ref{fig:folded_color}, which shows that
the color does vary over this time period, with the reddest emission
observed between phases $\sim$-0.5 to -0.2 while the bluest is at
phases $\sim$0 to 0.3. Possible origins for periodic variations in
color is described in this emission in \S\ref{sec:interpretation}.

\section{Discussion}
\label{sec:interpretation}

In this Section, we discuss possible physical origins for the main
results from the data analysis presented above:
\begin{enumerate}
    \item in \S\ref{sec:flare_disc} presenting an interpretation for
      the constant color (\S\ref{sec:color}), range in timescales
      (\S\ref{sec:flare}), and periodic nature
      (\S\ref{sec:periodicity}) of the long duration (multi-day)
      flares,  
    \item in \S\ref{sec:quiescent_disc} offer explanations for the
      color evolution (\S\ref{sec:color}) and periodicity of the
      ``quiescent'' emission, and 
    \item in \S\ref{sec:clump_disc}, the short time scale ($\sim$day)
      changes in the observed color (\S\ref{sec:color}) 
\end{enumerate}
we observe from IC 10 X-2.  This analysis is done within the context
of previous studies of this source, which as described in
\S\ref{sec:intro} found that IC 10 X$-$2 is a SFXT where a neutron
star (NS) \citep{Laycock_2014} is orbiting a massive star -- possibly
a LBV \citep{kwan_2018}.  In such a system, the optical and near-IR
emission studied is likely to be dominated by the massive star and/or
material accreting onto the NS.

Before calculating the properties of IC 10 X-2, it is necessary to
convert the apparent magnitudes measured in these observations to the
absolute magnitude $M_{\rm F}$ intrinsic to the source: 
\begin{eqnarray}
\label{eqn:mag}
M_{\rm F} & = & m_{\rm F}-A_{\rm F}-\mu
\end{eqnarray}
where $\mu$ is the distance modulus to IC 10 ($\mu \sim 24$;
\citealt{kim_2009, Laycock_2014}) and $A_F$ is the extinction along
the line of sight for a particular filter $F$.  The value of $A_{\rm
  F}$ depends on the total Hydrogen column density $N_{\rm H}$ towards
IC 10 X-2, the properties of the absorbing material -- which relates
$N_{\rm H}$ to $A_{\rm V}$, extinction in the visual $V$ band -- and
exctinction constant $y$ -- which relates $A_{\rm V}$ to $A_{\rm F}$.
From the X-ray spectrum of IC 10 X-2, \citet{Laycock_2014} measured
that a total $N_{\rm H}^{\rm tot}=6.3\times10^{21}~{\rm cm}^{-2}$,
which is combination of material in the Milky Way, in the foreground
interstellar medium of IC 10, and within IC 10 X-2 itself.  Galactic
H{\sc i} surveys suggest that, along this line of sight, within the
Milky Way (MW) $N_{\rm H}^{\rm MW} \approx 5.1\times10^{21}~{\rm
  cm}^{-2}$ \citep{hi4pi, kalberla05, dickey90}, and therefore
responsible for $\sim80\%$ of the total extinction observed towards
this source.  As a result, we estimate $A_{\rm V}^{\rm tot}$ using the
$N_{\rm H} - A_{\rm V}$ relation derived for MW dust  
\citep{Bohlin_1978, foight16, Zhu_2017, Whittet_2022}: 
\begin{eqnarray}
\label{eqn:NH}
A_{\rm V}^{\rm tot} & \approx & \frac{N_{\rm H}^{\rm
    tot}}{2.1\times10^{21}~{\rm cm}^{-2}}~{\rm mag} \approx 3~{\rm
  mag}. 
\end{eqnarray}
The extinction is a filter $A_{\rm F}$ is then: 
\begin{eqnarray} 
\label{eqn:A_lambda_y} 
A_{\rm F} = y_{\rm F} \times A_{\rm F} 
\end{eqnarray}
where $y_{\rm F}$ is the extinction constant provided by the SVO
Filter Service \citep{rodrigo12, rodrigo20}.  The values of $y_{\rm
  F}$, $A_{\rm F}$, and resultant absolute ``quiescent'' magnitude
$M_{\rm q,F}$ of IC 10 X-2 for all of the filters used in this work
are provided in Table \ref{tab:A_lambda}. 
 
\begin{table}[tb]
\caption{Extinction constant $y$, extinction magnitude $A_{\rm F}$,
  and quiescent absolute magnitude $M_{\rm q,F}$ of IC 10 X-2} 
\label{tab:A_lambda}
\begin{center}
\begin{tabular}{cccc} 
\hline\hline 
Filter & $y$ & $A_{\rm F}$ & $M_{\rm q,F}$ \\
$\cdots$ & $\cdots$ & mag & mag \\
\hline
$g^\prime$ & 1.21 & 3.63 & $-7.3$\\
$z_g$ & 1.19 & 3.57 & $-7.6$ \\
$r^\prime$ & 0.878 & 2.63 & $-7.5$\\
$z_r$ & 0.834 & 2.50 & $-7.8$ \\
$i^\prime$ & 0.669 & 2.00 & $-7.5$\\ \hline
\end{tabular}
\end{center}
\end{table}

\subsection{Nature of Long Duration Flares}
\label{sec:flare_disc}

As mentioned above, our analysis suggests that IC 10 X-2 periodically
(\S\ref{sec:periodicity}) produces flares ($m < m_{\rm flare}$) which
last $\gtrsim1 - 6$~ days (\S\ref{sec:flare}, Table
\ref{tab:flare_fit}) during which the color of this source is roughly
constant (\S\ref{sec:color}).  As discussed in \S\ref{sec:intro},
periodic flaring is a signature of BeHMXBs -- and therefore treat IC
10 X-2 as a member of this source class.  We note that this is not
inconsistent with the previous classifications of IC 10 X-2 as a SFXT
\citep{Laycock_2014} or having a LBV companion \citep{kwan_2018},
since both have been associated BeHMXBs (e.g., \citealt{Bird_2012,
  Clark_2013, Liu_2011}).  Furthermore, the absolute magnitude of IC
10 X-2 is similar to that of luminosity class Ia OeBe stars (e.g.,
\citealt{wegner06}), supporting this identification. 

In this scenario, the observed flares are result of the NS penetrating
the clumpy decretion disk produced by its supergiant stellar
companion, resulting in an increase in the accretion rate of material
onto the NS.  Therefore, the period of the flares is likely the
orbital period $P_{\rm orb}$ of the binary system that constitutes IC
10 X-2 $P_{\rm orb} \sim 26.5 - 26.6~{\rm d}$
(\S\ref{sec:periodicity}, Table \ref{tab:periodicity}) -- which is
very similar to the orbital period of other BeHMXBs \citep{fortin23},
such as LS~I+61$^\circ$303 ($P_{\rm orb} \approx 26.5~{\rm d}$;
\citealt{gregory02}).  Furthermore, the emission observed during these
flares is likely dominated by material accreting onto the NS.  The
expected hot temperature $T\sim10^6~{\rm K}$ of this material suggests
that, at the wavelengths of these observations (Table
\ref{tab:obsdata}), the observed emission would fall within the
Rayleigh-Jeans regime of the resultant blackbody radiation.  In this
case, the frequency dependence $\nu$ of the emitted flux density
$\S_\nu$ should be $S_\nu \propto \nu^{2}$, which correspond to an
unextincted color of: 
\begin{eqnarray}
    \label{eqn:ztf_color_bbody}
    (z_g-z_r)_0 & \approx & -0.64 \\
    \label{eqn:lco_color_bbody}
    (g^\prime - r^\prime)_0 & \approx & -0.57,
\end{eqnarray}
significantly bluer than the $(z_g-z_r)_{\rm obs} \approx 1.25$ and
$(g^\prime - r^\prime) \approx 1.1$ observed during these flares.  The
natural explanation for this discrepancy is that it is due to
extinction $A_v$ along the line of sight which, from Equations
\ref{eqn:mag} \& \ref{eqn:A_lambda_y}, is equal to: 
\begin{eqnarray}
    \label{eqn:flare_av_ztf}
    A_V^{z_g-z_r} & = & \frac{(z_g-z_r)_{\rm obs} -
      (z_g-z_r)_0}{y_{z_g}-y_{z_r}} \approx 5.31 \\ 
    \label{eqn:flare_av_lco}
    A_V^{g^\prime-r^\prime} & = & \frac{(g^\prime-r^\prime)_{\rm obs}
      - (g^\prime-r^\prime)_0}{y_{g^\prime}-y_{r^\prime}} \approx 5.03 
\end{eqnarray}
for the $y$ values given in Table \ref{tab:A_lambda} -- two magnitudes
of extinction more than inferred in Equation \ref{eqn:NH}.  This
suggests that, during the flares, the Hydrogen column density towards
material accreting on to the NS is $N_{\rm H}^{\rm flare} \sim
10^{22}~{\rm cm}^{-2}$.  This increase in $N_{\rm H}$ during the flare
could lead to a decrease and hardening of the X-ray emission produced
during this period, as observed previously from this source
\citep{Laycock_2014}, and similar changes and values of $N_H$ have
been observed in other HMXBs (e.g., IGR~J17544$-$2619;
\citealt{gonzalez04}).  A test of this hypothesis would be synchronous
optical and UV / X-ray observations of this source, to see if optical
flares are indeed associated with increases in extinction as observed
in other bands. 

Lastly, the duration of these flares would be connected to the
duration of the increased accretion onto the NS, which should occur
when it is within the decretion disk.  Therefore, the ratio of the
flare duration to orbital period provides an estimate of the fraction
of the NS's orbit spend inside this structure.  As listed in Table
\ref{tab:flare_fit}, most windows contain at least one flare
$\sigma\gtrsim2.5$~days long, suggesting that the decretion disk
encompasses $\gtrsim10\%$ of the NS's orbit.  This has implications
for the eccentricity of the NS's orbit, as well as the extent and
thickness of the decretion disk produced by the stellar companion,
which require additional information to disentangle. 

\subsection{Periodicity and Color Evolution of Quiescent Emission}
\label{sec:quiescent_disc}

As demonstrated above, our analysis of these observations of this
source suggests that, not only does the quiescent emission vary in
intensity on the same timescale as the orbital period $P_{\rm orb}$
(\S\ref{sec:periodicity}, Table \ref{tab:periodicity}), its color
changes as well -- with the quiescent emission being ``bluer'' when it
is brighter (\S\ref{sec:color}, Figure \ref{fig:CMDs}).  This is
opposite to what is observed in many other BeXRBs, where the  optical
emission is redder when brighter (e.g., Swift J0549.7-6812,
\citealt{Monageng_2019}; RX J0123.4-7321, \citealt{Coe_2021}; Swift
J010745.0-722740, \citealt{Schmidtke_2021}; RX J0529.8-6556
\citealt{Treiber_2021}; XMMU J010331.7-730144,
\citealt{Monageng_2020}; SXP 91.1, \citealt{Monageng_2019}; Swift
J010902.6-723710, \citealt{gaudin2024discovery}).  In these systems,
the optical reddening and increase in brightness are believed to
result in a scenario similar to that discussed in
\S\ref{sec:flare_disc} -- the passage of the NS through the decretion
disk leads to increased accretion, increasing its brightness, and
increased absorption along the line of sight, reddening the observed
emission.  Since we observe IC 10 X-2 to be ``bluer'' when brighter, a
different explanation is required.  Below, we discuss two possible
explanations for the color-magnitude behavior of IC 10 X-2: changes in
extinction through the decretion disk with orbital phase
(\S\ref{subsec:variable_nh}), and the formation of a transient
accretion disk around the NS (\S\ref{subsec:transient_disk}) as it
passes through the stellar decretion disk. 

\subsubsection{Orbital Variations in Extinction}
\label{subsec:variable_nh}

As detailed in \S\ref{sec:flare}, IC 10 X-2 is likely a BeHMXB where
the powerful stellar winds of the massive star companion generate a
decretion disk.  Depending on the geometry of the system, the line of
sight to the massive star passes through its decretion disk, which
will contribute to the total Hydrogen column density $N_{\rm H}$
towards the stellar companion.  The interaction between the neutron
star and decretion disk is expected to produce density inhomogeneities
in the disk (e.g., \citealt{manousakis15b}), which results in
variations in $N_{\rm H}$ through the disk over the orbital period of
the neutron star (e.g., \citealt{manousakis11, manousakis15}).  In
this model, we posit that the ``quiescent'' emission is dominated by
radiation from the stellar companion and that the observed changes in
brightness and color of this emission are due to changes in Hydrogen
column density $\Delta N_{\rm H}^{\rm orb}$ towards the star over the
orbit. 

To calculate $\Delta N_{\rm H}^{\rm orb}$ due to obscuring material
within IC 10 X-2, we first need to determine the conversion between
$N_{\rm H}$ and $A_{\rm F}$, which depends on the physical properties
of the internal material.  These quantities also determine the
constant $R$ relationship and the change in magnitude to that in color
due to the passage of light through the absorbing material:
  \begin{equation}
    \label{eqn:A_lambda_R} 
    A_{\lambda 1} = R_{\lambda 1} E(\lambda_{1}-\lambda_{2})
  \end{equation}
where $E(\lambda_{1}-\lambda_{2})$ is the reddening between
observations at wavelengths $\lambda_1$ and $\lambda_2$.  Since the
LCO data is significantly more precise than that collected by ZTF, we
use this dataset to calculate $R$.

\begin{figure}[tb]
    \centering
    \includegraphics[width=0.48\textwidth]{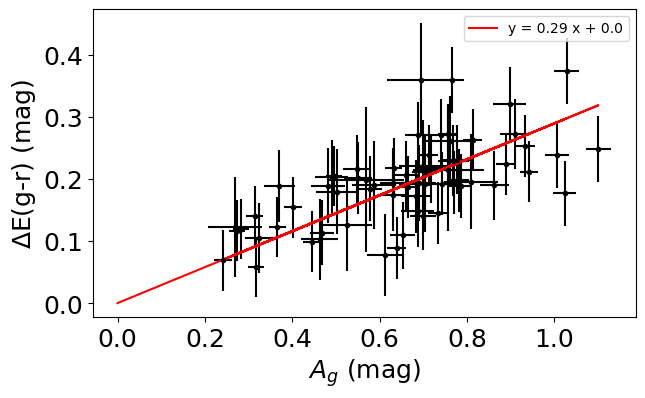}
    \caption{The color excess $\Delta E(g^\prime-r^\prime)$ against
      observed extinction $\Delta A_{g^\prime}$, with the best fit
      linear relation between these two parameters ({\it red line}).} 
    \label{fig:r_orbit}
\end{figure} 

In this scenario, the quiescent emission is bluest and brightest when
the extinction along the line of sight $A_{\rm F}$ is at a minimum.
From Figure \ref{fig:CMDs}, we estimate that, at this time (Figure
\ref{fig:CMDs}):
\begin{eqnarray*}
    m_{g^\prime}^{\rm bluest} & \sim & 19.54~{\rm mag},~{\rm and} \\
    (g^\prime-r^\prime)_{\rm bluest} & \sim & 1.02~{\rm mag}.
\end{eqnarray*}
Therefore, at a given time $t$, the change in extinction $\Delta
A_{g^\prime}(t)$ and reddening $\Delta E(g^{\prime} - r^{\prime}; t)$
towards the stellar component of IC 10 X-2 due to the intrasystem
material is: 
\begin{eqnarray}
\label{eqn:a_orbit}
    \Delta A_{g^\prime}(t) & = & m_{g^\prime}(t)-m_{g^\prime}^{\rm bluest} \\
\label{eqn:Eg-r}
    \Delta E(g^\prime-r^\prime; t) & = &
    (g^\prime-r^\prime)(t)-(g^\prime-r^\prime)_{\rm bluest}. 
\end{eqnarray}
As expected, $\Delta E(g^\prime - r^\prime)$ appears to increase for
larger values of $\Delta A_{g^\prime}$ (Figure \ref{fig:r_orbit}). We
use a least-squares-minimization routine to fit a line to these
parameters, with $R_{g^\prime}^{\rm orb}$ of the absorbing material
being the inverse of the resultant slope following Equation
\ref{eqn:A_lambda_R}).  This analysis suggest that the material inside
IC 10 X-2 has $R_{g^\prime}^{\rm orb} \sim 3.5$, similar to the value
measured for MW dust ($R_{g^\prime}^{\rm MW} \sim 3.3$;
\citealt{Yuan_2013}).  Therefore, the relation between $N_{\rm H}$ and
$A_{\rm V}$ derived for MW dust (Equation \ref{eqn:NH}) should be a
good approximation for the absorbing material within IC 10 X-2.

\begin{table}[bt]
\caption{Model Parameters (Equation \ref{eqn:sin_fit}) derived from
  fitting the folded ``quiescent'' ZTF light curve of IC 10 X-2} 
\label{tab:quiescent_sin_parameters}
\vspace*{-0.25cm}
\begin{center}
\begin{tabular}{ccc}
\hline
\hline
Parameter & $z_g$ & $z_r$ \\ 
\hline
A & $-0.16\pm0.01$ & $-0.11\pm0.0$ \\ 
$\phi_0$ & $0.06\pm0.01$ & $0.04\pm0.0$  \\ 
$m_q$ & $20.01\pm0.01$ & $18.82\pm0.0$ \\ 
$\chi^2$ / dof & 1337 / 303 & 2530 / 432 \\
 \hline
 \hline
\end{tabular}
\end{center}
\end{table}

The next step in calculating $\Delta N_{\rm H}^{\rm orb}$ is
determining the extinction $\Delta A_{\rm V}^{\rm orb}$ during an
orbit -- which in this interpretation is related to the amplitude $A$
of the periodic oscillations in the quiescent emission.  To do so, we
fit the folded $z_g$ and $z_r$ light curve with the same model used in
\S\ref{sec:periodicity} (Equation \ref{eqn:sin_fit}, but only using
data points with $m>m_{\rm flare}$ (Table \ref{tab:flare_threshold}),
which results in the parameters listed in Table
\ref{tab:quiescent_sin_parameters}.  We calculate $\Delta A_{\rm
  V}^{\rm orb,F}$ using the equation:
\begin{eqnarray}
    \label{eqn:av_orb}
    \Delta A_{\rm V}^{\rm orb,F} & = & \frac{2}{y_{\rm F}} |A_{\rm F}|,  
\end{eqnarray}
where $A_{\rm F}$ is the amplitude of the sinusoidal variations, and
$y_{\rm F}$ is the extinction coefficient (Table \ref{tab:A_lambda}),
for filter $F$.  For amplitudes $A$ measured from the $z_g$ and $z_r$
folded light curves given in Table \ref{tab:quiescent_sin_parameters},
we have:
\begin{eqnarray}
    \label{eqn:av_orb_zg}
    \Delta A_{\rm V}^{\rm orb, z_g} & \approx & 0.27~{\rm mag},~{\rm and} \\
    \label{eqn:av_orb_zr}
    \Delta A_{\rm V}^{\rm orb, z_r} & \approx & 0.26~{\rm mag}.
\end{eqnarray}
From Equation \ref{eqn:NH}, we therefore have:
\begin{eqnarray}
    \label{eqn:del_nh_orb}
    \Delta N_{\rm H}^{\rm orb} & \sim & 5.5\times10^{21}~{\rm cm}^{-2},
\end{eqnarray}
a quantity $\sim1/2$ of the $N_{\rm H}^{\rm flare}$ during the long
duration flares which occur when the neutron star is embedded in the
thickest part of the decretion disk (\S\ref{sec:flare_disc}).  This
value is smaller than inferred from other systems -- possibly the
result of differences in the stellar wind and observing geometry
(e.g., \citealt{manousakis11, manousakis15}).

\subsubsection{Transient accretion disk}
\label{subsec:transient_disk}

In this section, we investigate the possibility that the observed
periodic oscillations in the ``quiescent'' emission is the result of a
transient accretion disk around the neutron star.  As the neutron star
orbits its massive star companion, it is expected to accrete material
from its surroundings.  This material will likely be hotter, and
therefore its emission will be ``bluer,'' than the stellar companion.
Furthermore, the intensity of the emission is likely to vary over the
neutron star's orbit due to changes in the surrounding density, and
the timescale with which captured material falls onto the neutron
star.  When the intensity of emission from the disk grows, its
increased contribution to the total emission from the system will make
it appear brighter and bluer, as observed.  Therefore, a ``transient''
accretion disk -- as observed in other systems similar to IC 10 X-2
(e.g., \citealt{ducci10})-- could also explain the periodic changes in
brightness and color of the ``quiescent'' emission discussed in
\S\ref{sec:periodicity}.

Whether or not a transient accretion disk will form in a HMXB (e.g.,
\citealt{hainich20}) depends on the properties of the NS, its orbit,
and the wind produced by its SG companion.  As an initial
approximation of the orbital parameters, we assume the NS and SG are
moving in circular orbits of radius $d_{\rm NS}$ and $d_{\rm SG}$,
respectively, around their mutual center of mass, such that the total
distance $d_{\rm tot}$ between them:
\begin{equation}
\label{eqn:distance} 
d_{\rm tot}=d_{\rm NS}+d_{\rm SG}
\end{equation}
is constant with time.  In the simplifying case of a circular orbit, we have: 
\begin{eqnarray}
\label{eqn:r_cm} 
d_{\rm SG} & = & \frac{d_{\rm tot}\:M_{\rm NS} }{M_{\rm SG}+M_{\rm NS}} \\
\label{eqn:v_sg}
v_{\rm SG} & = & \sqrt{\frac{G\:M_{\rm NS}\:d_{\rm SG}}{d^{2}_{\rm tot}}} \\
\label{eqn:v_ns}
v_{\rm NS} & = &\sqrt{\frac{G\:M_{\rm SG}\:d_{\rm NS}}{d^{2}_{\rm tot}}} \\
\label{eqn:period}
P & = & \frac{2\pi\:d_{SG}}{v_{SG}}\\
\label{eqn:separation_distance2} 
d_{\rm tot} & = & \left(\frac{P}{2\pi}\right)^{\frac{2}{3}}[G(M_{\rm
    SG} + M_{\rm NS})]^{\frac{1}{3}}, 
\end{eqnarray} 
where $G$ is the gravitational constant.  For a SG mass $M_{\rm
  SG}=21.5~{\rm M}_\odot$ and NS mass $M_{\rm NS}=1.5~{\rm M}_\odot$,
as used by \citealt{kwan_2018}, this results in:  
\begin{eqnarray*}
d_{\rm tot} & \approx & 7.2\times10^{12}~{\rm cm}, \\ 
d_{\rm SG} & \approx & 5.1\times10^{11}~{\rm cm},~{\rm and} \\ 
d_{NS} & \approx & 6.7\times10^{12}~{\rm cm}. 
\end{eqnarray*}
We note that, given the typical radius of SG stars ($R_{\rm SG} \sim
10 ~R_\odot = 7\times10^{11}~{\rm cm}$; \citealt{Karino_2019}), the
center of mass of this system likely located near the edge of the
stellar companion.  

\begin{figure*}[]
\centering
    \subfigure{
    \includegraphics[width=0.9\textwidth]{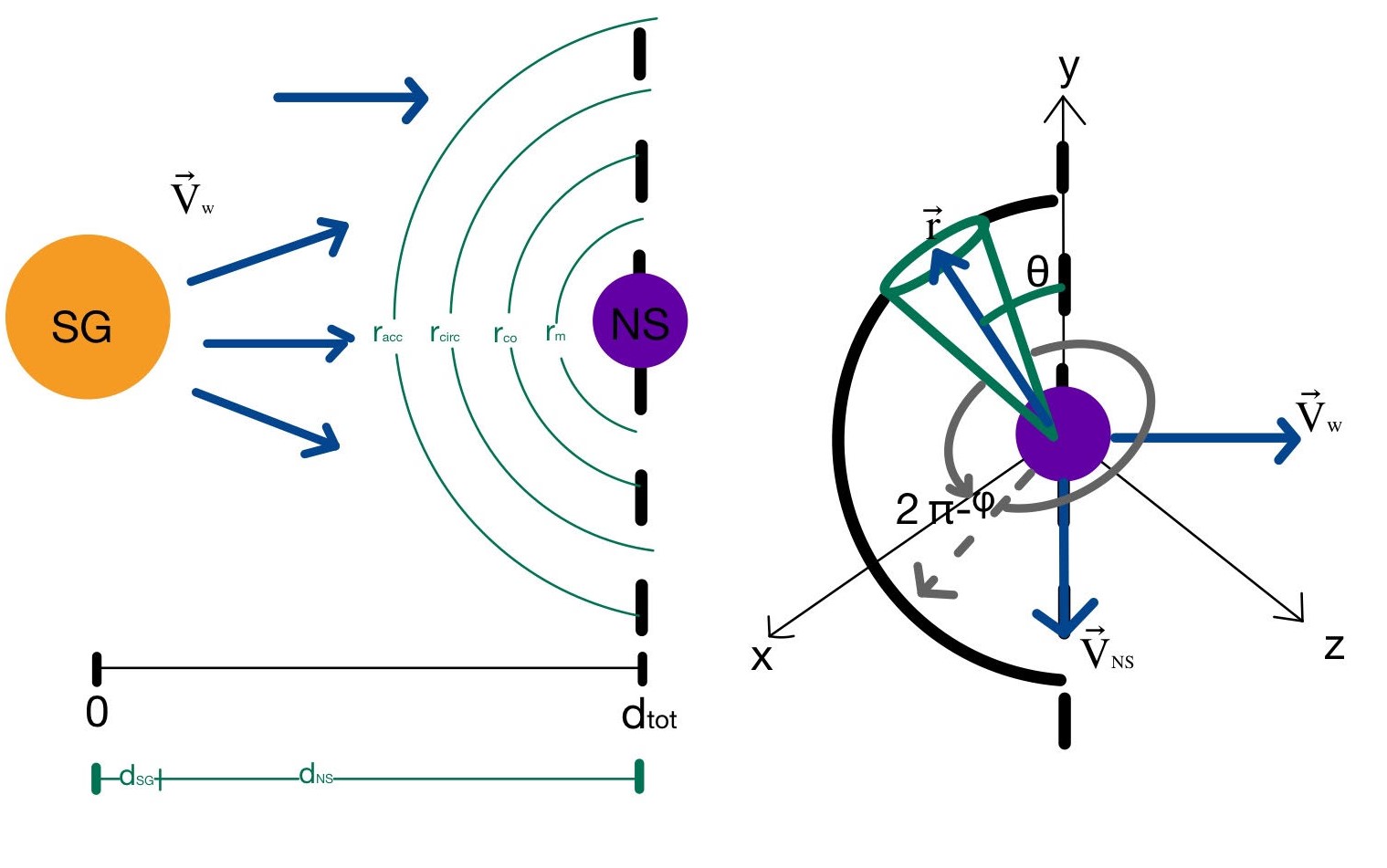}
    }    
    \caption{An illustration of IC 10 X-2 system geometry (see text
      for variables' definition)} 
    \label{fig:IC10X-2}
\end{figure*}

As shown in Figure \ref{fig:IC10X-2}, stellar wind material when
passes within a distance $r_{\rm acc}$ of the NS will be captured, and
eventually accreted, onto the compact object.  This distance is
approximated as where, relative to the NS, the potential and kinetic
energy of the stellar wind material are equivalent
\citep{Karino_2019}:  
\begin{eqnarray}
\label{eqn:r_acc}
\begin{split}
r_{\rm acc} & = \left(\frac{2\:G\:M_{\rm NS}}{v_{\rm rel}^{2}}\right) \\ 
       & = 4.7 \left(\frac{M_{\rm NS}}{1.5~M_\odot}\right)
\left(\frac{v_{\rm rel}}{900~\rm{km/s}} \right)^{-2} \;
\times10^{10}~\rm{cm} 
\end{split}
\end{eqnarray} 
where $G$ is the gravitational constant, $v_{\rm rel}$ is the relative
velocity, at the location of the NS $d_{\rm NS}$, between the stellar
wind $v_{\rm w}$ and the NS $v_{\rm ns}$.  Following work done by
\citet{Karino_2019}, we approximate $v_{\rm rel}$ as:  
\begin{equation}
\label{eqn:vrel_del} 
v_{\rm rel}^{2}=v_{\rm w}^{2}+v_{\rm NS}^{2}.
\end{equation} 
To determine the speed of the stellar wind $v_{\rm w}$, we use the
profile presented by \citealt{Karino_2019} among others, where:  
\begin{equation}
\label{eqn:v_w} 
v_{w}=v_{\infty}\left(1-\frac{R_{\rm SG}}{d_{\rm NS}}\right)^{\beta}.
\end{equation} 
For an acceleration parameter $\beta=1$, and a wind speed at
$r=\infty$ of $v_{\infty} = 10^{5}~{\rm km~s}^{-1}$ (e.g.,
\citealt{Karino_2019}), we find $r_{\rm acc} \approx
4.7\times10^{10}~{\rm cm}$ -- a small fraction of the distance between
the NS and SG.  

Stellar wind material entering within $r_{\rm acc}$ of the NS provides
an influx of angular momentum $J$ and mass $M$, which eventually
accretes on the NS.  For wind-fed X-ray binaries like IC 10 X-2,
whether or not this material forms a disk around the NS depends
requires that both (e.g., \citealt{Karino_2019}): 
\begin{enumerate}
    \item the circularization radius $r_{\rm circ}$, defined to be the
      distance from the NS where the initial angular momentum of the
      accreting material is equal to the angular momentum of a
      circular orbit around the NS, and  
    \item and the co-rotation radius $r_{\rm co}$, the distance from
      the NS where the period of a circular orbit is the same as its
      spin period, 
\end{enumerate}
 of the accreting material are larger than the neutron star's
 magnetospheric's radius $r_{\rm m}$, the distance from the NS where
 the accretion disk is disrupted by the NS's surface magnetic field
 \citep{Karino_2019}:  
\begin{equation}
\label{eqn:acc_disk_cond} 
r_{\rm m} < r_{\rm co}, r_{\rm circ}. 
\end{equation} 

As described by \citet{Karino_2019}, the circularization radius
$r_{\rm circ}$ is:  
\begin{equation}
\label{eqn:r_circ} 
r_{\rm circ}=\frac{l^2}{G \: M_{\rm NS}}
\end{equation} 
where $l$ is the specific angular momentum of stellar wind material
captured by the neutron star:  
\begin{equation}
l=\frac{\dot{J}}{\dot{M}}
\end{equation} 
where $\dot{J}$ and $\dot{M}$ are, respectively, the rate at which
angular momentum and mass from stellar wind material cross the
boundary $r=r_{\rm acc}$ from the NS (e.g.,
\citealt{Karino_2019}). Using the coordinate system defined in Figure
\ref{fig:IC10X-2}, these can be approximated as \citep{Karino_2019}:  
\begin{equation}
\label{eqn:j}
\begin{split}
\dot{J}& \approx r_{\rm acc}\: \rho_{w}\: v_{w}^{2} \int\limits_{0}^{\pi/2} 
\int\limits_{0}^{\pi} \sin^{2} \theta \cos\theta   \,  d\theta d\varphi \\
& \approx\frac{2}{3} \pi r_{acc}^{3}\: \rho_{w} \: v_{w}^{2}
\end{split}
\end{equation}
and
\begin{equation}
  \label{eqn:m_rate} 
\dot{M} \approx \dot{M_{w}}\:\frac{r^{2}_{acc}}{4\: d^{2}_{NS}}  
\end{equation}
where $\rho_{w}$ is the SG wind material density at $d_{NS}$.
Assuming conservation of mass flux, $\rho_w$ is approximately (e.g.,
\citealt{Karino_2019}):  
\begin{equation}
\label{eqn:rho}
    \rho_{w}=\frac{\dot{M_{w}}}{4 \pi d^{2}_{NS}\:v_{w}},
\end{equation} 
which results in: 
\begin{eqnarray}
    r_{\rm circ} & = & \frac{2}{3}\:\frac{r_{\rm acc}}{v_{\rm w}}\\
    & \sim & 4\times10^{10}~{\rm cm},
\end{eqnarray}
roughly $\sim80\%$ of $r_{\rm acc}$ calculated above.

Not surprisingly, The corotation radius depends on the unknown spin
period $P_{\rm spin}$ of the NS (e.g., \citealt{Karino_2019}): 
\begin{equation}
    \begin{split}
r_{\rm co} & = \left(\frac{ G \: M_{\rm NS} \: P_{\rm spin}^{2} }{
  4\pi^{2} }\right)^{\frac{1}{3}} \\  
      & \approx 3.7 \left(\frac{G\:M_{NS}}{4\pi^{2}\:1.5 \:
  M_\odot}\right)^{\frac{1}{3}} \left(\frac{P_{spin}}{100 \:
  \rm{s}}\right)^{\frac{2}{3}} \times10^{9} \; \rm{cm} 
    \end{split}
\end{equation}
which is currently unknown for the NS believed to be located in IC 10
X-2.  As shown in Table \ref{tab:trans_disk}, $r_{\rm co}< r_{\rm
  circ}$ even for extremely long spin periods  -- suggesting this
quantity will determine if a transient accretion disk is capable of
forming in IC 10 X-2. 

Lastly, we calculate the magenotospheric radius $r_{\rm m}$, which as
derived by \citet{Karino_2019} depends not only on the mass $M_{\rm
  NS}$, radius $R_{\rm NS}$, and surface magnetic field strength
$B_{\rm NS}$ of the neutron star, but also the mass loss rate of the
stellar companion $\dot{M}_{\rm w}$ and efficiency $xi$ of accretion
onto the NS (e.g., \citealt{Karino_2019}):  
\begin{equation}
\label{eqn:r_m} 
\begin{split}
r_{\rm m} & = \left( \frac{B_{\rm NS}^{4} R_{\rm NS}^{12} }{ 8\xi^{2}G
  M_{\rm NS} \dot{M}_{\rm w}^{2} }\right)^{\frac{1}{7}},\\  
& = 7.2 \left( \frac{B_{\rm NS}}{2\times10^{12}~\rm{G}}
\right)^{\frac{4}{7}} \left( \frac{R_{NS}}{10^{6}~\rm{cm}}
\right)^{\frac{12}{7}} \left( \frac{\xi}{0.5}
\right)^{-\frac{2}{7}}\\ 
&~~\left(\frac{M_{\rm NS}}{1.5~{\rm M}_\odot} \right)^{-\frac{1}{7}}
\left( \frac{\dot{M}_{\rm w}}{10^{-6}~\frac{\rm M_\odot}{\rm yr}}
\right)^{-\frac{2}{7}} \times 10^{7}~{\rm cm}. 
\end{split}
\end{equation}
While the range of possible values for most of these parameters is
small (less than an order of magnitude), $\dot{M}_{\rm w}$ ranges from
$\sim10^{-7}~{\rm M}_\odot~{\rm yr}^{-1}$ in Be stars (e.g.,
\citealt{lamers99}) to a ``typical'' value of $\sim10^{-6}~{\rm
  M}_\odot~{\rm yr}^{-1}$ in massive stars (e.g.,
\citealt{kudritzki00}) to $\sim10^{-4}~{\rm M}_\odot~{\rm yr}^{-1}$ in
red supergiants (e.g., \citealt{Mauron_2011}). However, as shown in
Table \ref{tab:trans_disk}, across this range $r_{\rm m}$ is
$\lesssim0.1r_{\rm co}$ -- suggesting that a transient accretion disk
can form in IC 10 X-2 per the criterion specified in Equation
\ref{eqn:acc_disk_cond}.  

\begin{table}[tb]
\caption{Physical properties of transient accretion disk, for
  different values of $\dot{M_w}$ and $P_{spin}$, assuming circular
  orbits.} 
\centering
\label{tab:trans_disk}
\begin{tabular}{llll}
\hline
\hline
$P_{\rm spin}$ [s] & 10 & 100 & 1000 \\
$r_{\rm co}$ [cm] & $8.0\times10^{8}$ & $3.7\times10^{9}$ &
$1.7\times10^{10}$ \\ 
\hline
$\dot{M}_{\rm w}$ $\left[ \frac{\rm M_\odot}{\rm yr}\right]$ 
 & $10^{-7}$ & $10^{-6}$ & $10^{-4}$ \\  
$\rho_{\rm w}$ $\left[\frac{\rm g}{\rm  cm^3}\right]$
&$1.2\times10^{-16}$ & $1.2\times10^{-15}$ & $1.2\times10^{-13}$ \\ 
$\dot{M}$ $\left[\frac{\rm M_\odot}{\rm  yr}\right]$ &
$1.2\times10^{-12}$ & $1.2\times10^{-11}$ & $1.2\times10^{-9}$ \\ 
$\dot{J}$ $\left[{\rm g}~\frac{\rm cm^2}{\rm s^2}\right]$ &
$2.2\times10^{32}$ & $2.2\times10^{33}$ & $2.2\times10^{35}$ \\ 
$r_{\rm m}$ [cm] & $1.4\times10^{-8}$ & $7.2\times10^{7}$ &
$1.90\times10^{7}$ \\ 
\hline
\hline
\end{tabular}
\end{table}

However, if a transient accretion disk is indeed responsible for the
periodic oscillations in the emission observed from IC 10 X-2, then
the luminosity of these variations $L_{\rm opt, disk}$ must be smaller
than the total radiative luminosity generated by accretion onto the NS
$L_{\rm acc}$: 
\begin{eqnarray}
    \label{eqn:lacc}
    L_{\rm acc} & \equiv & \eta \dot{M}c^2,
\end{eqnarray}
where $\eta$ is the radiative efficiency of the accreting matter.  The
amplitude of the oscillations derived from fitting the folded
quiescent light curve (Table \ref{tab:quiescent_sin_parameters})
suggests that, in the observed bands these oscillations have $\nu
L_\nu \sim 4\times10^{37}~{\rm ergs~s^{-1}}$.  While the relative
amplitude of the oscillation in quiescent emission observed in the
different filters studied in this work is consistent with the $L_\nu
\propto \nu^2$ expected from the Rayleigh-Jeans tail of a
multi-temperature blackbody radiation as expected from an accretion
disk heated by primarily viscous forces (e.g.,
\citealt{Shakura_1973,frank_king_raine_2002,russell_2008}), the errors
are sufficiently large to preclude modeling of this SED.  Therefore,
we approximate $L_{\rm opt,disk} \sim 4\times10^{37}~{\rm ergs
  s^{-1}}$.  Requiring $L_{\rm opt,disk} \leq L_{\rm acc}$ therefore
requires: 
\begin{eqnarray}
    \dot{M} & \geq & 8\left(\frac{\eta}{0.1}\right)^{-1}
    \times10^{-10}~\frac{\rm M_\odot}{\rm yr}. 
\end{eqnarray}
From the calculations above, assuming a circular orbit, this requires
the SG companion expels $\dot{M} \gtrsim 7.5 \times 10^{-5}~{\rm
  M_\odot~yr^{-1}}$ for $\eta \sim 0.1$ -- mass loss rates more
characteristic of red supergiants (e.g. \citealt{Mauron_2011}) than
the Be star (e.g., \citealt{lamers99, kudritzki00}) expected to reside
in IC 10 X-2. 

It is possible that, if the orbits in in IC 10 X-2 are eccentric -- as
observed from most BeHMXBs (e.g., \citealt{kretschmar19}) -- lower
values of $\dot{M}_{\rm w}$ are potentially possible.  If the NS
enters the decretion disk during an orbital phase near periapsis, the
NS will be closer to the SG, and moving faster, than in the case of a
circular orbit.  This is will increase both the velocity of the NS
relative to the stellar wind ($v_{\rm rel}$) and the density of the
stellar wind ($\rho_{\rm w}$) when the NS enters the decretion disk,
leading to an increase in $\dot{M}$ onto the NS.  However, it is
unclear the minimum stellar mass rate $\dot{M}_{\rm w}$ capable of
producing an accretion rate onto the NS $\dot{M}$ sufficient to power
the quiescent oscillations. 

\subsection{Short timescale color variability}
\label{sec:clump_disc}

As discussed in \S\ref{sec:color}, we also observed that, on several
occasions, that IC 10 X-2 gets significantly bluer or redder for a
short ($\lesssim1$ day) period of time (Figure \ref{fig:flares}).  The
short timescales of these event suggests the existence of small scale
structures -- either over dense or under dense regions -- in this
system.  There are multiple possible origins for clumps and holes
(e.g., \citealt{kretschmar19}) inside a sgHMXB such as IC 10 X-2
including instabilities in the stellar winds itself (e.g.,
\citealt{martinez-nunez17}) and the interaction between the NS and the
decretion disk (e.g., \citealt{manousakis15b, elmellah18}). 

One possible explanation for IC 10 X-2 getting ``redder'' is increased
absorption resulting from a dense clump passing through the line of
sight towards the main source of optical and infrared emission (which
is believed to be the SG star).  Following the analysis of
\S\ref{sec:flare_disc}, we can estimate the change in extinction
$\Delta A_V^{\rm clump}$ using the observed change in color during a
``reddening'' event (Equation \ref{eqn:flare_av_ztf}): 
\begin{eqnarray}
    \Delta A_V^{\rm clump} & = & \frac{(z_g-z_r)_{\rm obs} -
      (z_g-z_r)_0}{y_{z_g}-y_{z_r}}, 
\end{eqnarray}
where $(z_g-z_r)_0$ is the ``typical'' value of of $z_g-z_r$ for the
apparent magnitude $z_g$.  As shown in Figure \ref{fig:CMDs}, the most
extreme reddening event occurred had: 
\begin{eqnarray*}
    (z_g - z_r)_{\rm obs} & \approx & 2.25,~{\rm and} \\
    z_g & \approx & 20,
\end{eqnarray*}
where, for this value of $z_g$, typically IC 10 X-2 has $(z_g-z_r)_0
\sim 1.25$.  Therefore, during this event, $\Delta A_V^{\rm clump}
\sim 3$ -- an increase larger than inferred to occur during the long
duration flares (\S\ref{sec:flare_disc}), and comparable to the
average $A_{\rm V}^{\rm tot}$ towards IC 10 X-2 inferred from previous
X-ray observations (\S\ref{sec:interpretation},  Equation
\ref{eqn:NH}).  This suggests the obscuring clump responsible this
event had a Hydrogen column density $N_{\rm H}^{\rm clump} \sim
6\times10^{21}~{\rm cm}^{-2}$ -- comparable to what is inferred from
hydrodynamic simulations to exist in such systems (e.g.,
\citealt{elmellah18, manousakis15b}) and observations of similar
systems (e.g., \citealt{grinberg17}). 

Small scales ``holes'' in the distribution of material within this
system (e.g., \citealt{manousakis15b, elmellah18}) are unlikely to be
responisble for epochs when IC 10 X-2 is ``bluer'' than average for a
given magnitude since these structures are unlikely to significantly
decrease the total $N_{\rm H}$ along the line of sight (e.g.,
\citealt{grinberg17}).  We therefore consider the possibility they are
produced by a small, dense clump of material accreting onto the NS --
as invoked to explain short-term increases in the X-ray flux from such
sources (e.g., \citealt{manousakis15b}) -- whose resultant emission
dominates over that from the SG companion over the duration of the
accretion event.  This is the same physical scenario discussed in
\S\ref{sec:flare} for the long duration flares, and during such events
we expect emission in the observed bands to again be dominated by the
Rayleigh-Jeans regime of the multi-temperature blackbody radiation
generated by the accreting material, in which case the intrinsic
colors of IC 10 X-2 will be those given in Equations
\ref{eqn:ztf_color_bbody} \& \ref{eqn:lco_color_bbody}.  Assuming
$A_{\rm V}^{\rm tot}$ (Equation \ref{eqn:NH}) is the total extinction
towards the NS during this events, the observed $(z_g-z_r)_{\rm blue}$
color would be: 
\begin{eqnarray}
\label{eqn:ztf_color_blue}
    (z_g-z_r)_{\rm blue} & = & (y_{z_g}-y_{z_r})A_{\rm V}^{\rm tot} +
(z_g-z_r)_0 \\ 
    & \sim & 0.43. 
\end{eqnarray}
This value is comparable to the ``bluest'' emission observed from IC
10 X-2, $(z_g-z_r)_{\rm bluest} \sim 0.5$ (Figure \ref{fig:CMDs}),
suggesting that the decretion disk provides some additional absorption
during this time.  Since dense clumps are expected to be primarily
near the decretion disk (e.g., \citealt{grinberg17}), additional
absorption by the stellar wind material is not unexpected.  

\section{Conclusions}
\label{sec:conc}

In this work, we performed a comprehensive analysis of a regular
optical and near-infrared observations of IC 10 X-2 over a $\sim5$
year period with the ZTF and LCO facilities, which measured the
properties of this source in the $g^\prime$ / $z_{g}$, $r^\prime$
/$z_{r}$, and $i^\prime$ filters.  Our analysis indicates IC 10 X-2
exhibits both periodic flaring and oscillations in its ``quiescent''
emission (\S\ref{sec:periodicity}), suggesting it is a BeHMXB, a Be
star and neutron star orbiting each other with a $\sim26.5$ day period
-- a designation consistent with previous identifications as a SFXT
and sgHMXB (\S\ref{sec:interpretation}).  When the NS enters the
decretion disk generated by its stellar companion, the resultant
accretion of material onto the NS generates the observed multi-day
flares.  During these flaring episodes, emission is expected to be
dominated by the multi-temperature blackbody radiation emanating from
material in the accretion disk around the NS.  From the, nearly
constant, color observed during these flares (\S\ref{sec:flare}), we
infer that the column density towards IC 10 X-2 is $N_{\rm H}^{\rm
  flare} \sim 10^{22}~{\rm cm}^{-2}$ (\S\ref{sec:flare_disc}).  This
is $\sim1.5-2\times$ larger than the $N_{\rm H}^{\rm tot} \sim
6\times10^{21}~{\rm cm}^{-2}$ inferred from the X-ray spectrum of this
source (e.g., \citealt{Laycock_2014}; \S\ref{sec:interpretation}) --
the difference likely due to material in the decretion disk along the
line of sight.  For the observed periodic (\S\ref{sec:periodicity})
changes in color and brightness (\S\ref{sec:color}) of the
``quiescent'' emission of this source, we consider two possibilities
(\S\ref{sec:quiescent_disc}): orbital modulations in the column
density along the line of sight (\S\ref{subsec:variable_nh}) or the
varying contribution from a transient accretion disk around the NS
(\S\ref{subsec:transient_disk}).  We find that both explanations can
reproduce our results: the first requires that, during the orbit, the
$N_{\rm H}$ changes by $\Delta N_{\rm H}^{\rm orb} \sim
5.5\times10^{21}~{\rm cm}^{-2}$ -- comparable to the increase in
$N_{\rm H}$ observed during the long duration flares
(\S\ref{subsec:variable_nh}), while the second requires an accretion
rate $\dot{M} \gtrsim 8\times10^{-10}~{\rm M}_\odot~{\rm yr}^{-1}$ --
larger than expected for the typical mass-loss rate of a Be star
(\S\ref{subsec:transient_disk}).  Lastly, during the campaign we also
identified multiple epochs where the observed emisison from IC 10 X-2
was significantly redder and bluer than typically observed for its
apparent magnitude (\S\ref{sec:color} \& \S\ref{sec:flare}).  Since
these episodes were typically confined to one epoch, their short
durations ($\lesssim1$ day) argue for existence of small scale
structures, i.e. clumps, in material surrounding this binary system.
Epochs where IC 10 X-2 is significantly ``redder'' than expected is
likely to be the result of such clumps passing along the line of
sight, with the largest observed change in color suggesting the clump
had a column density $N_{\rm H}^{\rm clump} \sim 6\times10^{21}~{\rm
  cm}^{-2}$ (\S\ref{sec:clump_disc}).  Epochs where IC 10 X-2 is
significantly ``bluer'' than expected are possibly due to clumps
accreting onto the NS, with the resultant emission dominating
radiation in this band -- a conclusion supported by the similarity of
the bluest color observed by IC 10 X-2 and that expected from
blackbody radiation after accounting for the interstellar $N_{\rm H}$
along this line (\S\ref{sec:clump_disc}).   

Many of these conclusions can be tested by similar monitoring of the
UV and X-ray emission of this source, since increased accretion onto
the NS should lead to increase in emission in these wavebands.  If
that is correct, we expect that the X-ray flux of IC 10 X-2 would
increase, and possibly harden due to the increase $N_{\rm H}$, during
the observed optical flares.  Furthermore, if a transient accretion
disk is responsible for the observed periodic variations in the
quiescent emission, a similar periodicity may also be observed in the
X-ray band.  Further monitoring in the bands discussed in this work
(Table \ref{tab:obsdata}) would not only result in more examples of
the transient behaviors described in this work, but could also lead to
the discovery of longer timescale super-orbital variations (e.g.,
\citealt{kretschmar19}) -- providing additional information regarding
the interaction between the NS and its stellar companion in this
HMXB. 

\acknowledgments

JA was supported by the Kawader Research Program at NYU Abu Dhabi,
which is funded by the Executive Affairs Authority of the Emirate of
Abu Dhabi through the administration of Tamkeen. Both JA and JDG
received support form the NYU Abu Dhabi Research Institute grant to
the CASS.  COH is supported by NSERC Discovery Grant RGPIN-2016-04602.
This research has made use of the SVO Filter Profile Service "Carlos
Rodrigo", funded by MCIN/AEI/10.13039/501100011033/ through grant
PID2020-112949GB-I00.  This work has made use of the HMXB catalogue
(\url{https://binary-revolution.github.io/HMXBwebcat/}) maintained by
the Binary rEvolution team
(\url{https://github.com/Binary-rEvolution}).

\bibliography{Bibliographies}
\bibliographystyle{aasjournal}

\end{document}